\UseRawInputEncoding
\documentclass[journal=jpcbfk,manuscript=article]{achemso}
\usepackage{longtable} \usepackage{multirow} \usepackage{amsmath}
\usepackage{subfigure} \usepackage[usenames,dvipsnames]{color}
\usepackage{soul} \usepackage{threeparttable} 
\usepackage{float} \usepackage{bm}
\usepackage{subfig} \usepackage{amssymb}
\usepackage{pifont}
\usepackage[table]{xcolor}
\usepackage{xcolor}
\usepackage{multirow}

\usepackage{xr}

\makeatletter
\newcommand*{\addFileDependency}[1]{
  \typeout{(#1)}
  \@addtofilelist{#1}
  \IfFileExists{#1}{}{\typeout{No file #1.}}
}
\makeatother

\newcommand*{\myexternaldocument}[1]{%
    \externaldocument{#1}%
    \addFileDependency{#1.tex}%
    \addFileDependency{#1.aux}%
}

\myexternaldocument{supplement}

\author{Seyedeh Maryam Salehi, Silvan K\"aser, Kai T\"opfer}
\affiliation[University of Basel]{Department of Chemistry, University
  of Basel, Klingelbergstrasse 80 , CH-4056 Basel, Switzerland.}
\author{Polydefkis Diamantis} \affiliation[EPFL] {Laboratory of
  Computational Chemistry and Biochemistry, Institute of Chemical
  Sciences and Engineering, \'{E}cole Polytechnique F\'{e}d\'{e}rale
  de Lausanne, CH-1015 Lausanne, Switzerland.}  \author{Rolf Pfister,
  Peter Hamm} \affiliation[University of Zurich] {Department of
  Chemistry, University of Zurich} \author{Ursula R\"othlisberger}
\affiliation[EPFL] {Laboratory of Computational Chemistry and
  Biochemistry, Institute of Chemical Sciences and Engineering,
  \'{E}cole Polytechnique F\'{e}d\'{e}rale de Lausanne (EPFL), CH-1015
  Lausanne, Switzerland.}  \author{Markus Meuwly}
\affiliation[University of Basel]{Department of Chemistry, University
  of Basel, Klingelbergstrasse 80 , CH-4056 Basel, Switzerland.}
\email{m.meuwly@unibas.ch}

\title{Hydration Dynamics and IR Spectroscopy of 4-Fluorophenol}

\begin{document}

\begin{abstract}
Halogenated groups are relevant in pharmaceutical applications and
potentially useful spectroscopic probes for infrared spectroscopy. In
this work, the structural dynamics and infrared spectroscopy of
$para$-fluorophenol (F-PhOH) and phenol (PhOH) is investigated in the
gas phase and in water using a combination of experiment and molecular
dynamics (MD) simulations. The gas phase and solvent dynamics around
F-PhOH and PhOH is characterized from atomistic simulations using
empirical energy functions with point charges or multipoles for the
electrostatics, Machine-Learning (ML) based parametrization and with
full \textit{ab initio} (QM) and mixed Quantum Mechanical/Molecular
Mechanics (QM/MM) simulations with a particular focus on the CF- and
OH-stretch region. The CF-stretch band is heavily mixed with other
modes whereas the OH-stretch in solution displays a characteristic
high-frequency peak around 3600 cm$^{-1}$ most likely associated with
the -OH group of PhOH and F-PhOH together with a characteristic
progression below 3000 cm$^{-1}$ due to coupling with water modes
which is also reproduced by several of the simulations. Solvent and
radial distribution functions indicate that the CF-site is largely
hydrophobic except for simulations using point charges which renders
them unsuited for correctly describing hydration and dynamics around
fluorinated sites.
\end{abstract}

\section{Introduction}
Fluorination - and halogenation in general - are common chemical
modifications for pharmaceuticals. Approximately 20 \% of all small
molecule drugs used in medicinal chemistry contain ${\rm X} = {\rm
  F}$, Cl, Br, or I or a combination thereof. Among these compounds
halogenated phenyl rings constitute an important
class.\cite{halog-biotech2013meyer} Because of the directionality of
the interaction along the C-X bond due to the sigma hole, halogenation
has emerged as one of the essential chemical modifications in
medicinal
materials\cite{hernandes2010halogen,matter2009pi-binding,Muller2007fluorine},
and supramolecular
chemistry.\cite{metrangolo2005supra,metrangolo2008supra} By changing
the halogen atom, the interactions with the environment can be tuned
and the hydrophobicity around the modification site can be
modulated.\cite{lommerse1996halogens,Auffinger2004halobonds,Hobza2011design,diederich2011halogens,matter2009pi-binding,riley2011tuning1,Elhage2015janus}
The importance of halogenation as a fundamental concept in medicinal
chemistry is highlighted by the improved binding affinities of several
ligands towards their
receptors.\cite{Lu2009rationalDD,Wilcken2013application} Recently,
halogenation has also been employed in the context of protein
modifications, such as for insulin, to fine-tune thermodynamic
stability and affinity to the insulin
receptor.\cite{MM.insulin:2016}\\

\noindent
A halogen bond ``[..]occurs when there is evidence of a net attractive
interaction between an electrophilic region associated with a halogen
atom in a molecular entity and a nucleophilic region in another, or
the same, molecular entity.''\cite{Desiraju.hbond:2013} Hence, halogen
atoms act as electrophiles and can form an attractive interaction with
a nucleophilic counterpart. Based on the analysis of the molecular
surface electrostatic potential (ESP),\cite{clark.sigmahole:2007} the
``halogen bond'' was also associated with a ``$\sigma$-hole
bond''\cite{rev.sigmahole:2016} which is a noncovalent interaction
between a covalently-bonded halogen atom X and a negative site, e.g. a
lone pair of a Lewis base or an anion.\cite{clark.sigmahole:2007} Such
a ``bond'' involves a region of positive electrostatic potential,
i.e. the $\sigma$-hole, and as an extension of one of the covalent
bonds to the atom. The $\sigma$-hole arises as a consequence of the
anisotropy of the ESP around the halogen atom. The strengths of the
interactions generally correlate well with the magnitudes of the
positive and negative electrostatic potentials of the $\sigma$-hole
and the negative site. As fluorine has the largest electronegativity
and the lowest polarizability, for some time it was in fact assumed
that there is no $\sigma$-hole and that therefore fluorine is not
involved in halogen bonding at all.\cite{clark.sigmahole:2007,
  Neaton:2017, meyer:2015} However it is now well established that it
can have a positive $\sigma-$hole and form halogen bonds when it is
linked to strongly electron-withdrawing groups including Cl, Br, and
I.\cite{politzer.sigmahole:2011,clark.sigmahole:2013} Moreover, there
is also experimental evidence for fluorine engaging in halogen
bonding.\cite{legon:1999}\\

\noindent
Introducing a fluorine atom into organic molecules can cause major
changes in the physico-chemical properties such as solubility,
chemical reactivity and biological activity compared to
non-fluorinated analogues.\cite{shah:2007} In particular, fluorine
often replaces hydrogen in organic molecules but the size and
stereoelectronic influences of the two atoms (hydrogen vs. fluorine)
are quite different albeit it is often regarded as isosteric
substitution.\cite{Muller2007fluorine} In bio-inorganic and medicinal
chemistry, the formation of intermolecular O--H/F--C and N--H/F--C
hydrogen bridges was assumed to be important in binding fluorinated
compounds to enzyme active sites.\cite{barbarich:1999} Such
interactions affect enzyme ligand binding affinity, selectively
coupled with the changes in pharmaco-kinetic properties by fluorine
substitution.\cite{diederich2011halogens,shah:2007} The effects of
fluorine substitution on the related pharmaco-kinetic properties like
lipophilicity, volatility, solubility, hydrogen bonding and steric
effects affect the resulting compound binding, absorption, transport
and hence the related biological activity.\cite{fluorine:2011}\\

\noindent
A variety of functional groups, including C-H, C-OH, C=O, and
C$\equiv$N, have utilized the C-F bond as a
bioisostere.\cite{Rachel:2020} However, it is difficult to generalize
the relative ability of fluorine to act similar to a hydrogen or
hydroxy group, and different factors must be considered in each
case. The van der Waals radius of fluorine (1.47 {\AA}) lies between
that of oxygen (1.57 {\AA}) and hydrogen (1.2 {\AA}) and as it is the
element with the highest electronegativity, the C-F bond is almost
identical to C-OH in terms of bond length and polarity. Despite its
three electron pairs, the C-F bond interacts more weakly with the
environment compared to an oxygen atom and is better described as
``weakly polar'' rather than ``hydrogen
bonding''.\cite{fluorine:1997,Rachel:2020} In pharmacological
applications the replacement H$\rightarrow$F is often considered to
avoid metabolic transformation due to the high stability of the CF
bond. Examples are drugs interacting with P450 for which fluorination
has been widely used to block metabolic
transformations.\cite{Rachel:2020}\\

\noindent
Given the different qualitative characterizations outlined so far, a
more molecularly refined picture of the energetics and dynamics
involving fluorinated model compounds will be valuable. The present
work considers hydrated fluoro-phenol (F-PhOH) as a typical
representative. Using linear infrared spectroscopy together with
computational characterizations at different levels of theory the
structural dynamics and spectroscopy of F-PhOH is characterized. The
computations use advanced empirical force fields including multipolar
interactions, a machine-learned, neural network-based representation
of a full-dimensional PES, mixed quantum mechanics/molecular mechanics
and {\it ab initio} MD simulation techniques. The measured data from
infrared spectroscopy can be directly compared with the computational
results. Also, the solvent dynamics is investigated based on frequency
fluctuation correlation and spatial distribution functions. In this
article, first, the methods are described followed by the results for
the spectroscopy of the -CF and -OH from experiments and
simulations. Then, the solvent structure is analyzed from radial
distribution functions and from 2-dimensional solvent distributions
and finally, conclusions are drawn.\\

\section{Methods}
\subsection{Classical Molecular Dynamics Simulations} 
All classical MD simulations were performed with
CHARMM\cite{Brooks.charmm:2009}. The bonded parameters are based on
CGenFF \cite{CGenFF:2010} except for the CF and OH bond for which a
Morse potential was used to describe their anharmonicity. To that end,
a scan along the CF bond was performed at the MP2/aug-cc-pVTZ level
starting from an optimized structure of F-PhOH at this level of
theory. The energy of 49 points was computed on a grid ranging from
$r=0.75$ \AA\/ to $r=5.55$ \AA\/ in increments of 0.1 \AA\/. Then, the
energies were fitted to a Morse potential $V(r) =
D_{0}[1-\exp(-\beta(r-r_{0}))]^{2}$ which yields parameters $D_{0} =
136.316$ kcal/mol, $r_{0} = 1.349$ \AA\/, and $\beta = 1.603$
\AA\/$^{-1}$. For the OH bond the calculated Morse parameters are
$D_{0} = 120.234$ kcal/mol, $r_{0} = 0.971$ \AA\/, and $\beta = 2.088$
\AA\/$^{-1}$. To realistically describe the electrostatic
interactions, a multipolar
(MTP)\cite{Kramer2012,MM.mtp:2013,MM.mtp:2016,MM.chroma:2017} model
was also used with MTPs on all heavy atoms up to quadrupoles and point
charges for all hydrogen atoms. These parameters were fitted to the
electrostatic potential using a fitting
environment\cite{mm.mtp2:2016}, see Tables S1 to S3.\\

\noindent
Simulations of F-PhOH and PhOH were carried out in a cubic box of
$30^3$ \AA\/$^3$ ($28^3$ \AA\/$^3$ for simulations with the NN-PES,
see below) using TIP3P\cite{Jorgensen.tip3p:1983} water
molecules. Minimization, heating, and equilibration procedures for 40
ps were employed to prepare the system. Production simulations of 5 ns
were run in the $NVE$ ensemble at 300 K using a Velocity
Verlet\cite{Swope.vv:1982}. The time step was $\Delta t = 1$ fs and
every 5 snapshot was recorded. Lennard-Jones interactions were
computed with a 12 \AA\/ cutoff switched at 10
\AA\/.\cite{Steinbach.cutoff:1994} The electrostatic interactions for
the monopoles (point charges) are treated using Particle-Mesh
Ewald\cite{Darden.pme:1993} (PME) with grid size spacing of 1 \AA\/,
characteristic reciprocal length $\kappa = 0.32$ \AA\/$^{-1}$, and
interpolation order 6. All bonds involving hydrogen atoms are
constrained via the SHAKE
algorithm.\cite{shake77,Gunsteren.shake:1997} Additional MD
simulations were also carried out for PhOH in water with the same
setup that was used for F-PhOH in order to directly compare their
spectroscopy and solvent structure.\\

\noindent
For the simulations with the NN-based PES (see below) the atomic
simulation environment (ASE) was used.\cite{larsen2017atomic} The
van-der-Waals interactions were those from the
CGenFF\cite{CGenFF:2010} parametrization and the fluctuating charges
are from the PhysNet representation, see below. For both terms
interactions the cutoff distance is at 14~\AA\/ and switched between
13 to 14~\AA\/. To avoid artifacts of the electrostatic Coulomb force
in the cutoff range, the Coulomb force at the distance of 14~\AA\/ was
shifted to zero in accordance to the shifted forces
method.\cite{Spohr1997} In the gas phase 1000 trajectories, each
200~ps in length, are run to obtain an ensemble average. The $NVE$
simulations are run at 300~K initialized from random momenta
corresponding to a Maxwell-Boltzmann distribution, with a time step of
0.5~fs, equilibrated for 50~ps and propagated for 200~ps. Simulations
in solution $NVT$ using Langevin\cite{Langevin_MD} thermostat at 300~K
are performed for 20 trajectories of 100~ps each with a time step of
0.2~fs to obtain a total of 2~ns for PhOH and F-PhOH in solution,
respectively. The IR spectra are then calculated from the
dipole-dipole moment autocorrelation function
\cite{TRAVIS3,schmitz2004vibrational,schmitz2004vibrational2} and
averaged over all 1000 trajectories.\\

\subsection{Instantaneous Normal Mode Analysis}
From the production simulation, $10^6$ snapshots were taken as a
time-ordered series for computing the frequency fluctuation
correlation function (FFCF). The FFCF was determined from
instantaneous harmonic vibrational frequencies based on a normal mode
analysis. Such instantaneous normal modes (INM) are obtained by
minimizing F-PhOH while keeping the surrounding solvent frozen. Next,
normal modes were calculated using CHARMM for 5 modes ($\nu_1$ to
$\nu_5$ in ascending order) between 1100 to 1400 cm$^{-1}$ in terms of
participation ratio of CF stretch in that particular mode. In a
separate analysis step, the participation ratios of the CF, CO, and CH
stretch and the COH bending coordinates to these 5 normal modes were
determined.\\

\subsection{Frequency Fluctuation Correlation Function and Lineshape}
From the INMs the frequency trajectory $\omega_i(t)$ and the FFCF,
$\langle \delta \omega(0) \delta \omega(t) \rangle$ was
computed. Here, $\delta \omega(t) = \omega(t) - \langle \omega(t)
\rangle$ and $\langle \omega(t) \rangle$ is the ensemble average of
the transition frequency. From the FFCF the line shape function
\begin{equation}
\label{eq:1}
g(t) = \int_{0}^{t} \int_{0}^{\tau^{'}} \langle \delta
\omega(\tau^{''}) \delta \omega(0) \rangle d\tau^{''} d\tau^{'}.
\end{equation}
is determined within the cumulant approximation. To compute $g(t)$,
the FFCF is numerically integrated using the trapezoidal rule and the
1D-IR spectrum is calculated from\cite{2DIRbook-Hamm-2011}
\begin{equation}
\label{eq:2}
I(\omega) = 2 \Re \int^\infty_0
e^{i(\omega-\langle\omega\rangle)t} e^{-g(t)} e^{-\frac{t \alpha}{2T_1}}dt
\end{equation}
where $\langle\omega\rangle$ is the average transition frequency
obtained from the distribution, $T_1 =1.2$ ps\cite{MM.facn:2015} is
the vibrational relaxation time and $\alpha = 0.5$ is a
phenomenological factor to account for lifetime
broadening.\cite{2DIRbook-Hamm-2011}\\

\noindent
From the FFCF, the decay time is determined by fitting the FFCF to a
general expression\cite{hynes:2004}
\begin{equation}
  \label{eq:ffcffit}
  \langle \delta \omega(t) \delta \omega(0) \rangle = \sum_{i=2}^{n}
  a_{i} e^{-t/\tau_{i}} + \Delta_0
\end{equation}
where $a_{i}$, $\tau_{i}$ and $\Delta_0$ are fitting parameters. The
decay times $\tau_i$ from the fits characterize the time scale of the
solvent fluctuations. The absence of a minimum at short times ($\tau
\sim 0.02$ ps) indicates that the interaction between F and
environment is weak compare with situation in F-ACN or
N$_3^{-}$.\cite{MM.facn:2015,MM.n3:2019} The decay times $\tau_i$ of
the FFCF reflect the characteristic time-scale of the solvent
fluctuations to which the solute degrees of freedom are coupled. In
all cases the FFCFs were fitted to an expression containing two decay
times using an automated curve fitting tool from the SciPy
library.\cite{2020SciPy-NMeth}

\subsection{Full \textit{Ab Initio} (QM) and Mixed Quantum Mechanical/Molecular Mechanics (QM/MM) Simulations}
{\bf Full QM Simulations:} The QM system was comprised of F-PhOH and
117 water molecules in a (15.41 \AA\/ ,15.44 \AA\/ ,15.46 \AA\/)
periodic box initially equilibrated classically at 300 K and 1 atm
using CHARMM. The full QM equilibration and production phases lasted
for 12.5 ps and 20.4 ps respectively. For the gas phase simulation,
the total equilibration and production times were 23.0 ps and 28.1 ps,
respectively. For the latter, the initial translations and rotations
of the center of mass were removed.\\

\noindent
For both the gas phase and the condensed phase systems, the full QM
simulation protocol consisted of (i) an equilibration of the system at
300 K first with Born-Oppenheimer (BO) MD and then with Car-Parrinello
(CP) MD \cite{PhysRevLett.55.2471}, and (ii) a production phase in the
microcanonical ($NVE$) ensemble. The respective time steps for BO and
CP MD were 10 and 2 atomic units (a.u.), respectively.  In CP MD, the
fictitious electron mass was equal to 400 a.u. In the production
phase, frames were saved every 10 a.u., corresponding to a time
interval of approximately 0.48 fs. \\

\noindent
Density Functional Theory (DFT)-based \textit{ab initio} MD
simulations of F-PhOH in gas phase and in aqueous solution were
carried out using the CPMD code \cite{CPMD} using the BLYP functional
for the exchange and correlation energies\cite{blyp:1988,parr:1988}
with the addition of Dispersion-Corrected Atom-Centered Potentials
(DCACPs)\cite{PhysRevLett.93.153004,PhysRevB.75.205131,JCTC.9.955} for
the description of dispersion forces.\\ Norm-conserving
Martins-Trouiller pseudopotentials\cite{troullier:1991} were used in
combination with a plane wave basis with a 175 Rydberg kinetic energy
cutoff of for the expansion of the single-particle wavefunctions. The
latter value was selected because it reproduces a converged
equilibrium C-F bond distance at the BLYP-DCACP level of 1.35 \AA\/
for F-PhOH in gas phase, which is in good agreement with values
obtained at the MP2/6-311++G(df,pd) (1.34 \AA\/) and
B3LYP/6-311++G(df,pd) (1.35 \AA\/) levels,
respectively.\cite{Michalska.IRpfoh:2003}\\

\noindent
{\bf Mixed QM/MM Simulations}: Two QM/MM MD simulations were carried
out for F-PhOH and PhOH in water, respectively, using the QM/MM
interface of CPMD with the Gromos code\cite{GROMOS} and the coupling
scheme developed by Rothlisberger and
coworkers.\cite{QMMM1,QMMM2,QMMM3} The two systems were comprised of
the solute (F-PhOH or PhOH), and 331 and 311 water molecules,
respectively. The system size was selected so that a direct comparison
with the full QM simulation of F-PhOH in water can be made, and assess
the impact of quantum description of the solvent including
polarization effects on the geometric and spectral properties of
F-PhOH. \\

\noindent 
The systems were first equilibrated classically, using
AMBER18.\cite{AMBER} F-PhOH and PhOH were modelled with the GAFF2
force field \cite{GAFF,GAFF2}, while the TIP3P model was used for
water. Following an initial minimization, the two systems were
equilibrated in the isothermal-isobaric $(NPT)$ (300 K, 1 atm)
ensemble with the Berendsen barostat\cite{Ber_bar} and Langevin
dynamics\cite{Langevin_MD} for pressure and temperature control
respectively, followed by simulations in the NVT ensemble with
Langevin dynamics, for a total of 100 ns. A time step of 2 fs was
employed. In view of the small periodic box size, a reduced real space
cutoff of 7 \AA\/ was used for the nonbonded interactions. \\

\noindent
In both systems, the solute was treated at the QM level and the
solvent at the classical (MM) level. The QM setup, and the simulation
time step were the same as described above for the full QM
simulations. The QM/MM MD simulation protocol was also similar to the
one described for the full QM simulation, apart from the use of two
separate Nos\'e-Hoover thermostats for the QM and MM parts
respectively, during the equilibration with BO and CP MD. For the
F-PhOH system, the equilibration and production runs lasted 10.1 ps
and 35.9 ps respectively, while for the PhOH system they lasted 12.6
ps and 25.0 ps, respectively. During the production phase, frames were
saved with the same frequency as in the full QM simulations (0.48
fs).\\

\subsection{Machine-Learned Potential Energy Surface (PES)}
To validate in particular the PC- and MTP-based simulations using an
empirical force field a complementary model based on a machine-learned
PES was also pursued. For this PhysNet\cite{MM.physnet:2019}, a deep
neural network (NN) of the message passing
type\cite{gilmer2017neural}, was used to obtain a representation of
the potential energy for both PhOH and F-PhOH. PhysNet uses Cartesian
coordinates and nuclear charges to learn an atomic descriptor for the
prediction of energies, forces, partial charges and molecular dipole
moments to describe chemical systems and their properties, such as
infrared spectra.\\

\noindent
PhysNet was trained on \textit{ab initio} energies, forces and dipole
moments calculated at the MP2/6-31G(d,p) level of theory using
Molpro\cite{molpro19} according to the protocol reported in
Ref.~\cite{MM.physnet:2019}. The reference data, containing different
geometries for both molecules, is generated from MD simulations at 50,
300 and 1000~K using CHARMM force field (5000 geometries each yielding
a total of 30000 geometries) and extended with geometries obtained
from normal mode sampling\cite{smith2017ani} at temperatures between
10 and 2000~K (6600 geometries for each molecule). The complete data
set thus contains 43200 PhOH and F-PhOH structures. The performance of
the PhysNet PES is reported in Figure S1 which shows
the correlation between the reference MP2 and the PhysNet energies for
a test set of 3700 randomly selected points with $R^2 = 0.9999$ and an
RMSE of 0.0037 eV. Simulations with PhysNet for the solute and an
empirical water model are subsequently referred to as ML/MM MD.\\

\subsection{Experimental Infrared Spectroscopy}
Experimental spectra of PhOH and F-PhOH in water have been measured in
attenuated total reflection (ATR) geometry using a home-constructed
ZnSe prism in a Bruker Tensor 27 FT-IR spectrometer. Identical
concentrations of 0.76~M have been used for both samples, which is the
saturation limit for F-PhOH. As a background spectrum, pure water has
been measured as well and subtracted from the PhOH and F-PhOH
spectra. The subtraction procedure is deemed reliable in the shown
spectra, with the exception of the range between $\approx
3100$~cm$^{-1}$ to 3500~cm$^{-1}$, where the very strong OH stretch
vibration dominates. In fact, in this range a dispersion-shaped
response is observed, which is attributed to the effect the samples
have on the water spectrum, i.e., a shift to somewhat higher
frequencies due to overall weaker water-water hydrogen bonds.

\section*{Results}
\subsection{Gas Phase Spectra}
First, the performance of the PC- and MTP-based empirical force
fields, of PhysNet, and of the DFT-BLYP/DCACP based {\it ab initio} MD
simulations was assessed for F-PhOH in the gas phase. For this, MD
simulations of F-PhOH in the gas phase were carried out and the power
and infrared spectra were determined and compared with experiments.\\

\begin{figure}[H]
\begin{center}
\includegraphics[width=0.95\textwidth]{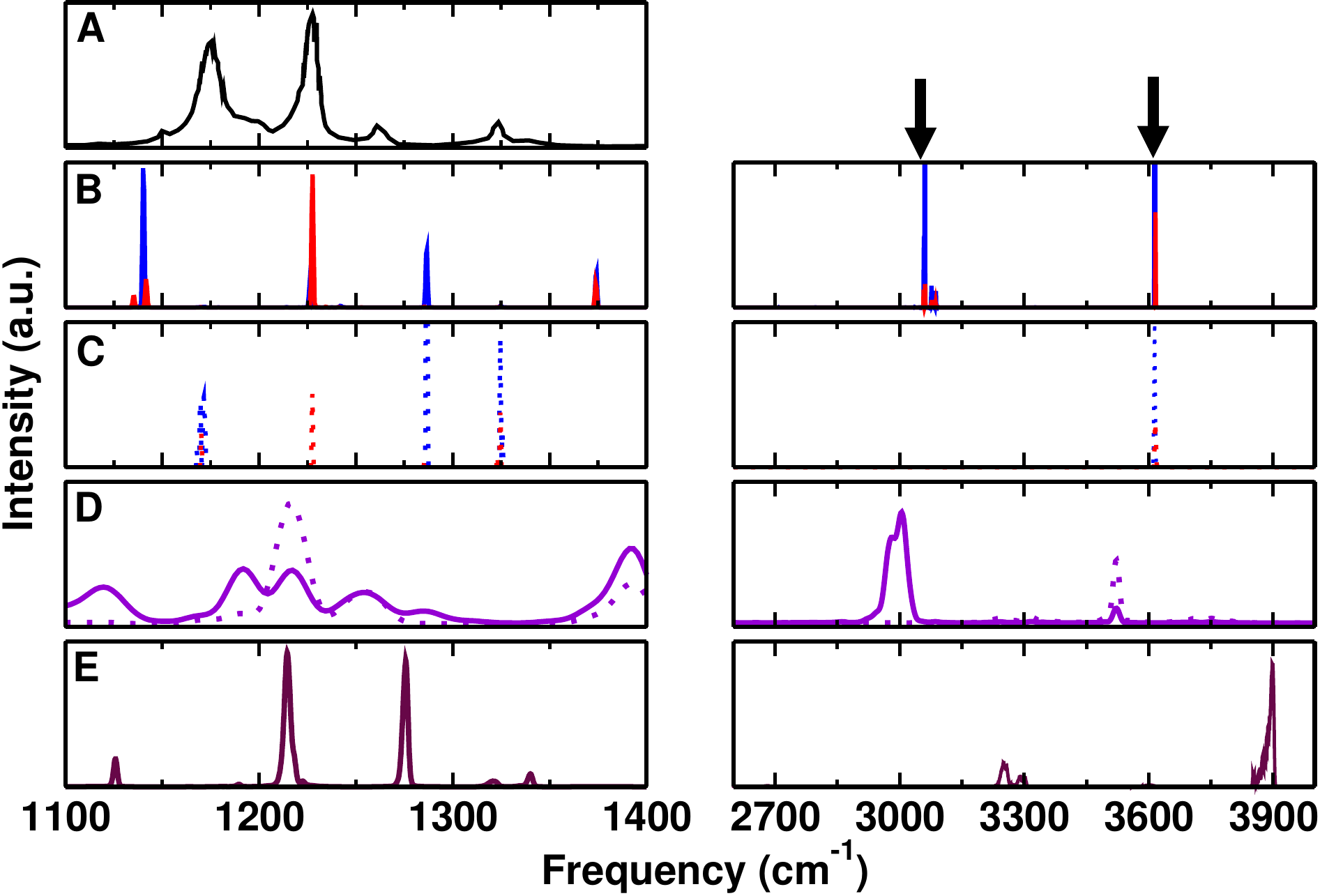}
\caption{Comparison of experimental (in CCl$_4$) and computed (in the
  gas phase) spectra for F-PhOH in the range of (1100-1400 cm$^{-1}$)
  and (2600 to 4000 cm$^{-1}$). Panel A: experimental (black) spectrum
  in CCl$_4$ from Ref. \cite{Michalska.IRpfoh:2003} extracted using
  g3data.\cite{g3data} The arrows in the right hand column refer to
  the experimental frequencies for the CH and OH
  stretches.\cite{Michalska.IRpfoh:2003} Panels B and C: IR from the
  Fourier transform of the total dipole moment correlation function
  and CF/OH power spectra from PC (blue) and MTP (red) simulations for
  F-PhOH in the gas phase. Panel D: global (solid) and CF/OH (dotted)
  power spectrum in violet from QM simulations in the gas phase. Panel
  E: IR spectrum (maroon) obtained from ML/MM MD simulations.}
\label{fig:spec_gas}
\end{center}
\end{figure}

\noindent
Figure \ref{fig:spec_gas} reports the infrared and CF-/OH-power
spectra of F-PhOH from simulations with the experimental FT-IR
spectrum\cite{Michalska.IRpfoh:2003} from 1100 to 1400 cm$^{-1}$ and
between 2600 and 4000 cm$^{-1}$, respectively. For clarity, the left
hand column shows the low-frequency vibrations whereas the right hand
column is for the -OH-stretch region. The experiments were carried out
in CCl$_4$ solvent and the spectral lines have a full width at half
maximum of $\sim 10$ cm$^{-1}$. As an indication for the
solvent-induced shift incurred, for PhOH in CCl$_4$ the CO stretch is
found at 1257 cm$^{-1}$ which amounts to a red shift of $\sim -5$
cm$^{-1}$ compared with the gas-phase frequency of 1261.7
cm$^{-1}$.\cite{Michalska.IRphenol:2001,Bist.irphenol:1967} Hence,
CCl$_4$-induced shifts for F-PhOH are expected to be a few wavenumbers
as well. To the best of our knowledge, no gas-phase spectra are
available for F-PhOH. Hence, the frequencies for F-PhOH measured in
CCl$_4$ are used {\it in lieu} of gas phase data.\\

\noindent
The measured CF and CO stretching modes occur mainly at 1226 and 1262
cm$^{-1}$, respectively, while they couple to one another and
potentially to other modes. According to the
analysis\cite{Michalska.IRpfoh:2003, Bist.irphenol:1967} (see Table
\ref{tab:irexpgas}), the CF stretch is coupled to the in plane bending
of the ring and also the C-H bend while the CO stretch couples to the
CC and CF stretching vibrations. Therefore, the CF stretch is
intimately coupled with other modes and for that reason it is not
possible to assign a local CF stretching mode to one particular
frequency.\\

\begin{table}[H]
\caption{Vibrational frequencies in cm$^{-1}$ for PhOH (gas phase) and
  F-PhOH (in CCl$_4$) in the range of 1100-1400
  cm$^{-1}$.\cite{Bist.irphenol:1967,Michalska.IRphenol:2001,Michalska.IRpfoh:2003}
  The contributions (in terms of local deformations) to each
  vibrational mode indicate strong mixing and are those from the
  literature.\cite{Michalska.IRphenol:2001,Michalska.IRpfoh:2003} The
  assignment of the bands has been made on the basis of the calculated
  potential energy
  distribution.\cite{morino:1952,Michalska.IRpfoh:2003} Symbols $\nu$
  and $\delta$ refer to stretching and bending modes, respectively.}
\begin{tabular}{l|cccccccccccc}
 \hline\hline
  PhOH      &1150.7       &1168.9       &1176.5       &               &1261.7       &1343  & &3656\\        
  \hline
            &$\delta$(CH) &$\delta$(CH) &$\delta$(OH) &               &$\nu$(CO)    &$\delta$(CH) & &$\nu$(OH) \\
            & $\nu$(CC)   &$\nu$(CC)    &$\delta$(CH) &               &$\delta$(CH) &$\delta$(OH) \\
            & $\delta$(OH)&             &$\nu$(CC)    &               &             &                        \\
            \hline
  F-PhOH    &1149         &1174        &              &1226           &1262         &1310&1323 & 3613 \\
  \hline
            &$\delta$(CH)&$\delta$(OH) &              &$\nu$(CF)     &$\nu$(CO)     &$\delta$(CH)&$\nu$(CC) &$\nu$(OH) \\
            &            &$\nu$(CC)    &              &$\delta $ring  &$\nu$(CC)    &$\nu$(CC)   &$\delta$(OH) \\
            &            &$\delta$(CH) &              &$\delta$(CH)  &$\nu$(CF)     &            &$\delta$(CH) \\
 \hline\hline
\end{tabular}
\label{tab:irexpgas}
\end{table}

\noindent
For the force field simulations with PC and MTP the $\beta-$parameter
of the CF-Morse potential (see Methods) was slightly adjusted to
$\beta = 1.665$ \AA\/$^{-1}$ to correctly describe the experimental
spectrum in CCl$_4$ (Figure \ref{fig:spec_gas}A), hence the favourable
comparison with the IR spectra in Figure \ref{fig:spec_gas}B. The CF
power spectrum (Figure \ref{fig:spec_gas}C left) clarifies that this
mode couples strongly to other vibrations close in frequency for both,
PC (blue) and MTP (red) force fields. The peak structure from the
CF-power spectrum in Figure \ref{fig:spec_gas}C left matches that from
the experiment whereas for the IR spectrum in panel B this is only
qualitatively the case. For the QM simulations in the gas phase
(Figure \ref{fig:spec_gas}D) the two main peaks of QM are at 1217 and
1256 cm$^{-1}$ and appear to be shifted by 5 to 6 cm$^{-1}$ with
respect to experiments at 1222 and 1262 cm$^{-1}$ which were recorded
at $T=300$ K. Finally, MD simulations using the PhysNet representation
of the MP2/6-31G(d,p) reference data the IR-spectrum in Figure
\ref{fig:spec_gas}E shows two strong peaks at 1215 cm$^{-1}$ and 1276
cm$^{-1}$.\\

\noindent
In the region of the OH-stretch vibration, the experimental spectrum
for F-PhOH in CCl$_4$ reports a band at 3613 cm$^{-1}$ (black
arrow).\cite{Michalska.IRpfoh:2003} This compares with 3614 cm$^{-1}$
and 3616 cm$^{-1}$ from PC and MTP simulations, respectively, Using
PhysNet, the main peak in the gas phase is at 3889 cm$^{-1}$ (harmonic
frequency at 3882 cm$^{-1}$ at the MP2/6-31G(d,p) level; corrected
frequency at 3655 cm$^{-1}$ by multiplying with 0.94 for this level of
theory\cite{Radom-scale:1996}) whereas the DFT-BLYP/DCACP QM
simulations report the OH stretch at 3522 cm$^{-1}$, somewhat shifted
to the blue and red, respectively, compared with experiment.
Moreover, there are signatures in the infrared spectra due to the
CH-stretch vibrations (black arrow) around 3035-3077 cm$^{-1}$ which
are also observed in the MD simulations. For PhysNet the corresponding
peak is at 3256 cm$^{-1}$ (3060 cm$^{-1}$ after correction with a
scaling factor of 0.94) whereas for QM at the DFT-BLYP/DCACP level the
absorption is at 3006 cm$^{-1}$. The power spectra from the PC and MTP
simulations (Figure \ref{fig:spec_gas}C) confirm that the OH-stretch
is a local mode.\\

\noindent
In summary, the gas phase spectrum from finite-$T$ MD simulations find
comparable patterns for the frequencies in the 1100 to 1400 cm$^{-1}$
region when compared with experiment. It is also found that the CF
stretch is coupled to other modes in this spectral range and no local
mode for this motion can be assigned.\\

\subsection{Spectroscopy and Dynamics of F-PhOH in Water}
\begin{figure}[H]
\begin{center}
  \includegraphics[width=0.95\textwidth]{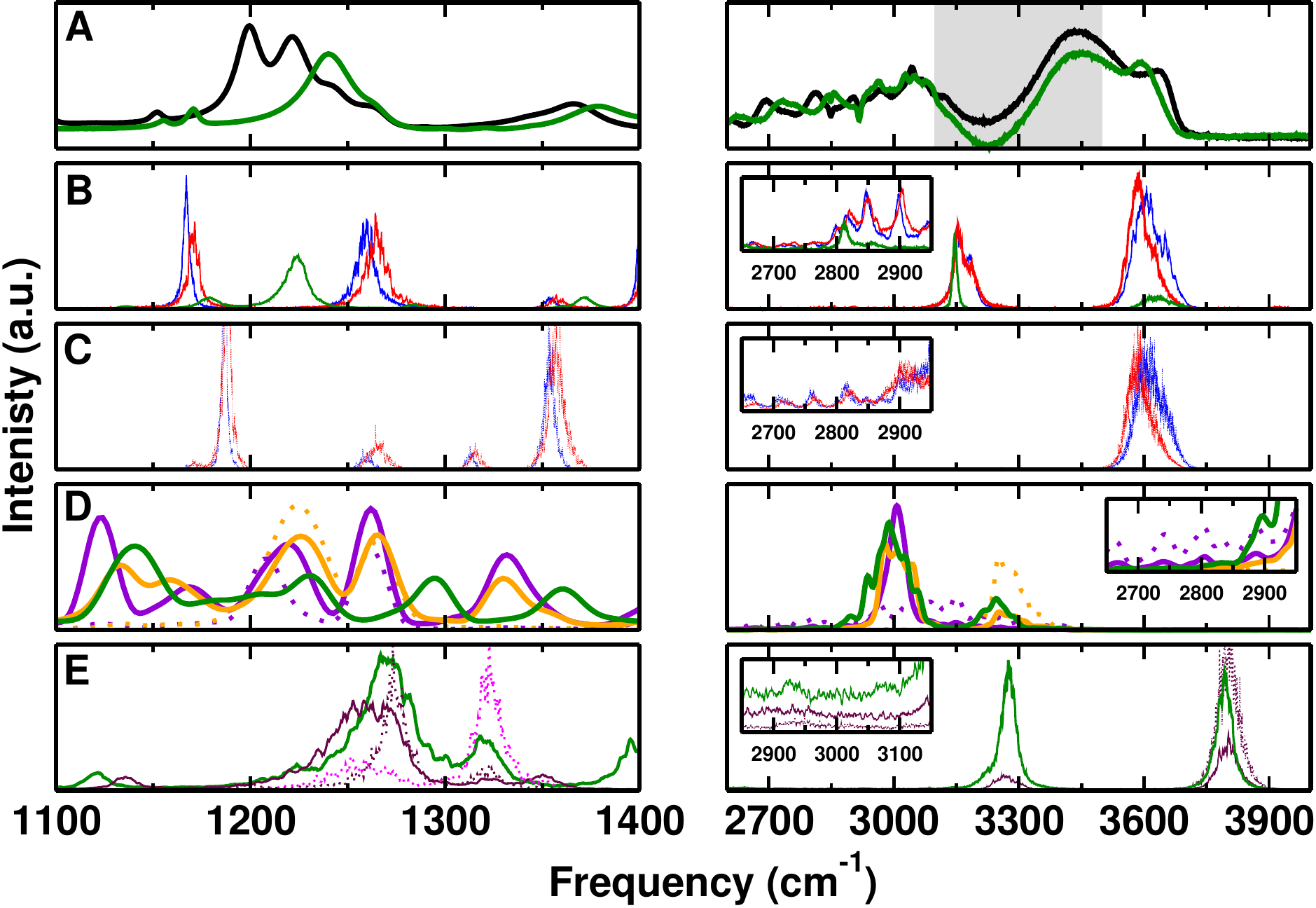}
\caption{Comparison of experimental and computed spectra for F-PhOH
  and PhOH in solution for the ranges (1100-1400 cm$^{-1}$) and (2600
  to 4000 cm$^{-1}$). Panel A: Measured spectrum for F-PhOH (black)
  and PhOH (green). The gray area from 3100~cm$^{-1}$ to
  3500~cm$^{-1}$ is unreliable due to incomplete subtraction of the
  water background. Panel B: IR spectra for F-PhOH (PC (blue) and MTP
  (red)) and PhOH (MTP (green)). Panel C: CF/OH power spectra for
  F-PhOH from PC (blue) and MTP (red) simulations. Panel D: Global
  (solid) and CF/OH (dotted) power spectrum of the solute only
  (dotted) for F-PhOH (QM (violet) and QM/MM (orange)) and PhOH (QM/MM
  (green)). Panel E: IR (solid) and CF/OH (dotted) power spectrum for
  F-PhOH (maroon) and PhOH (green) from ML/MM simulations. The magenta
  dotted line is the CO power spectrum of F-PhOH. Insets in panels B
  to D show that the structured spectrum below 3000 cm$^{-1}$ is also
  found from the simulations whereas with PhysNet this is less
  realistically captured compared with experiment.}
\label{fig:spec_solv}
\end{center}
\end{figure}

After assessing the energy functions considered in the present work,
the spectroscopy of F-PhOH in solution is analyzed, see Figure
\ref{fig:spec_solv}. The experimentally measured spectrum from the
present work is the black trace in Figure \ref{fig:spec_solv}A with
two prominent bands at 1201 and 1222 cm$^{-1}$ in the CF-stretch
region together with additional unresolved shoulders to higher energy,
superimposed on a broad background extending from 1170 to 1270
cm$^{-1}$.\\

\noindent
The MTP/MD simulations for the CF power spectrum (Panel C) show two
prominent peaks at 1187, and 1264 cm$^{-1}$ which approximately line
up with the features in the infrared spectrum (Panel B) but are
displaced from those observed experimentally. The peak at 1171
cm$^{-1}$ is red shifted compared to the double peak at 1201 and 1222
cm$^{-1}$ of experimental spectrum while the peak at 1264 cm$^{-1}$ is
blue shifted or captured at the same position compared to two
additional experimental peaks at 1244 and 1264 cm$^{-1}$. Furthermore,
the smaller peak at 1357 cm$^{-1}$ is also red shifted compared to
1368 cm$^{-1}$ from experiment. The corresponding infrared spectra in
panel B align with the features at 1260 cm$^{-1}$ but the signal at
1187 cm$^{-1}$ from the power spectra has no oscillator strength in
the infrared spectrum. Instead, a peak in the infrared appears at 1171
cm$^{-1}.$\\

\noindent
The {\it ab initio} MD simulations in solution for the CF-power
spectrum (Figure \ref{fig:spec_solv}D solid violet) find two prominent
bands at 1207 and 1262 cm$^{-1}$ compared with band maxima at 1217
cm$^{-1}$ with faint shoulders below 1200 cm$^{-1}$ from gas-phase
simulations, see Figure \ref{fig:spec_gas}D. The two prominent bands
are also found from QM/MM simulations (solid orange) with band maxima
at 1225 and 1266 cm$^{-1}$. The CF-power spectra assign these features
to CF-stretch involving motions and accounting for the scaling for
BLYP calculations ($\sim 0.99$ largely independent of basis
set)\cite{Radom-scale:1996} they shift to the red which is consistent
with the experiments. Thus, the splitting between the two peaks
decreases from 55 cm$^{-1}$ for the full QM simulations to 41
cm$^{-1}$ indicating the sensitivity of the solvent interactions and
the resulting frequency shifts and splittings in this spectral
range.\\

\noindent
For ML/MM MD simulations a broad band is observed between 1220 to 1290
cm$^{-1}$ in the IR spectrum which is blue shifted compared to the
experiment. The CF-power spectrum with peak maximum at 1273 cm$^{-1}$
(dotted maroon trace in Figure \ref{fig:spec_gas}E) indicates that
part of this broad IR-lineshape is due to CF-stretching
motion. Moreover, the CO power spectrum has a first peak at 1250
cm$^{-1}$ which contributes to the broad IR peak below 1300 cm$^{-1}$
and a second, prominent peak at 1323 cm$^{-1}$, see Figure
\ref{fig:spec_gas}E. The overlapping peaks of the CF- and CO-power
spectra indicate coupling between the two types of motion.\\

\noindent
The high-frequency region of the experimental spectrum for solvated
F-PhOH (Figure \ref{fig:spec_solv}A) above and below 3000 cm$^{-1}$
involves a broad absorption extending from $\sim 2700$ to 3100
cm$^{-1}$, a region between 3100 and 3500 cm$^{-1}$ that is not
reliable due to the dominating background from the OH stretch
vibration of bulk water which can not be subtracted off completely
(grey area in Figure \ref{fig:spec_solv}A), and a high-frequency
feature at 3643 cm$^{-1}$ assigned to the free OH vibration presumably
originating from F-PhOH but possibly also from H$_2$O. This latter
assignment is less likely, though, as the same sharp peak appears for
PhOH in solution but shifted to the red by $\sim 40$ cm$^{-1}$. If the
signal was due to water it is expected to occur at closer frequencies
given the similarity of the solutes. Furthermore, experiments on
hydrated PhOH with up to 49 water molecules find the water-OH-stretch
vibration at $\sim3700$ cm$^{-1}$.\cite{mizuse:2009} From simulations
with PCs and MTPs the high frequency peak at 3607 cm$^{-1}$ is
consistent with experiment. Comparison with the spectrum for F-PhOH in
CCl$_4$ and PhOH in the gas phase shows that this signal corresponds
to the ``free OH stretch'' vibration, see Figure
\ref{fig:spec_gas}. Features at 3150 cm$^{-1}$ in the infrared
spectrum (Figure \ref{fig:spec_gas}B) are due to the CH-stretch
vibrations which could be brought into better agreement with
experiment by slight reparametrization of the force constants. These
features are not present in the OH-power spectra, Figure
\ref{fig:spec_gas}C, as expected which confirms the assignment to the
CH-stretch vibration. The structured spectrum below 3000 cm$^{-1}$ is
present in both, the PC and MTP simulations, albeit with lower
intensity.\\

\noindent
For the full QM and QM/MM simulations the global power spectra (Figure
\ref{fig:spec_solv}D solid violet and orange) find a signal centered
at 3000 cm$^{-1}$ which is typical for the CH-stretch modes. At higher
frequency ($\sim 3250$ cm$^{-1}$) the OH-stretch vibration is located
which is confirmed by the OH-power spectra (dotted violet and orange
traces). However, no signal in the 3600 cm$^{-1}$ region is present
which suggests that the ``free OH'' signature in these simulations,
expected around 3500 cm$^{-1}$ from the QM gas phase simulations
(Figure \ref{fig:spec_gas}D), is absent.\\

\noindent
Simulations with the PhysNet energy function primarily find the high
frequency -OH stretch at 3803 cm$^{-1}$ with broad, largely
unstructured undulations below 3000 cm$^{-1}$. It is likely that the
MP2/6-31G(d,p) level is not sufficient for quantitatively describing
the spectroscopy of F-PhOH. Accounting for a frequency scaling of 0.94
for harmonic frequencies\cite{Radom-scale:1996} shifts all frequencies
to the red which is more consistent with the experimentally determined
spectra. Specifically, the 3800 cm$^{-1}$ and 3280 cm$^{-1}$ band
maxima shift to 3572 cm$^{-1}$ and 3083 cm$^{-1}$, both of which are
consistent with OH- and CH-stretching motions.\\

\noindent
The broad feature below 3000 cm$^{-1}$ from the experiments deserves
additional attention. Regular signatures in this frequency range were
previously reported for thin film liquid PhOH and solid
PhOH\cite{evans:1960} and for PhOH at the air/water
interface.\cite{tahara:2018} Such regular structures have been
observed also in other hydrogen-bonded systems, such as the acetic
acid dimer, and are typically used to characterize a medium-strong
hydrogen bond.\cite{nibbering:2007} They are attributed to a
Franck-Condon-like progression of the hydrogen-bond vibration (with a
frequency of ca. 50 cm$^{-1}$ in the present case) that is
anharmonically coupled to the high-frequency OH stretch vibration.\\

\noindent
Simulations for F-PhOH in solution with PC/TIP3P, MTP/TIP3P, and full
QM show an extended spectroscopic response in this frequency range
with pronounced peaks superimposed which are washed out in the
ML/TIP3P simulation and entirely absent in the QM/MM simulations. This
suggests that the spectroscopic signature below 3000 cm$^{-1}$ is due
to coupling between the H-bonding motion of water around the -COH part
of F-PhOH which is primarily sensitive to the nonbonded
interactions. To assess whether or not flexibility of the water
solvent also affects the spectral signatures, simulations with the
reparametrized,\cite{Burnham97p6192,MM.ice:2008} flexible KKY
(Kumagai, Kawamura, Yokokawa) model were carried out.\cite{kky_orig}
One 5 ns simulation with a time step of $\Delta t = 0.25$ fs for
F-PhOH was run and analyzed. The power spectrum of the F-PhOH
OH-stretch vibration confirms the pronounced, regular pattern with
peaks separated by some $\sim 50$ cm$^{-1}$ below 3000 cm$^{-1}$, see
Figure S2. In addition, the main peak between 3300
cm$^{-1}$ and 3600 cm$^{-1}$ shifts to the red by 78 cm$^{-1}$
compared with simulations using the rigid TIP3P water model. This
confirms that the pattern below 3000 cm$^{-1}$ is due to anharmonic
coupling through nonbonded interactions between solute and solvent and
not caused by the water internal modes.\\

\begin{figure}[H]
\begin{center}
\includegraphics[width=0.4\textwidth, angle=-90]{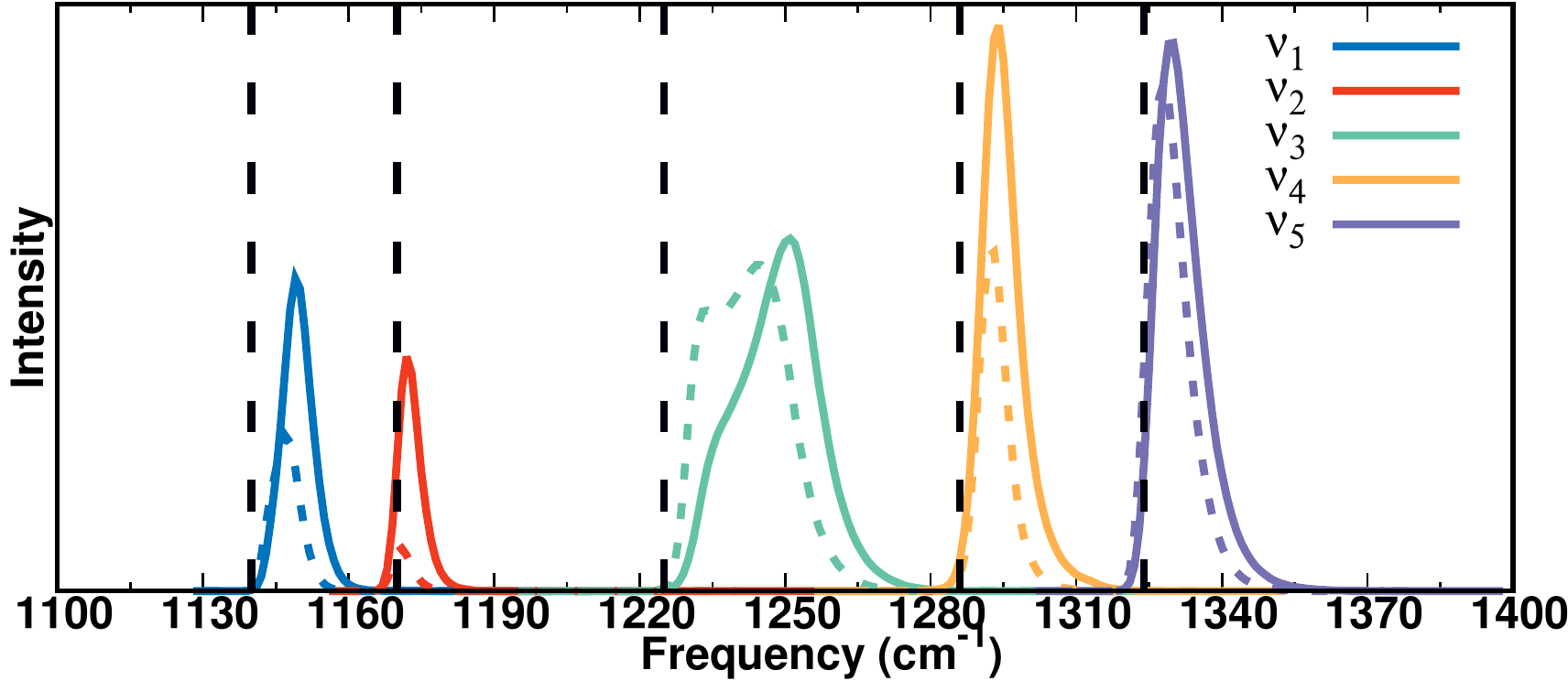}
\caption{Instantaneous vibrational frequency distributions from 5 ns
  MTP (solid colored line) and PC (dashed colored line) simulations of
  F-PhOH in water for five modes between 1100 and 1400 cm$^{-1}$. The
  black dashed lines are the harmonic frequencies for the optimized
  structure in the gas phase using the PC model.}
\label{fig:inmpcmtp}
\end{center}
\end{figure}

\noindent
For assessing solvent-induced frequency shifts, the frequency
distributions from 5 ns simulation of hydrated F-PhOH, analyzed with
instantaneous normal modes (INM) for the PC and MTP model are compared
with the normal modes from gas phase simulations, see Figures
\ref{fig:inmpcmtp} and/or Table S4 for the frequency
maxima. For obtaining the instantaneous normal modes the water
environment was frozen and the structure of the solute was optimized,
followed by a normal mode calculation. The band positions compare well
with the frequencies from Table \ref{tab:irexpgas}. However, the
bimodal distribution around 1250 cm$^{-1}$ for both, simulations with
PC and MTP, can not be convincingly correlated with the experimental
spectra.\\

\noindent
The participation ratios (see Methods) from the MTP simulations of the
local modes to the frequencies of the five modes in the 1100 to 1400
cm$^{-1}$ range ($\nu_1$ to $\nu_5$) are shown in Figure
S3. These were determined from the normal modes of
F-PhOH over $10^5$ snapshots in solution. The contributions of the CF,
CO, and CH stretch and the COH bending modes to each of the vibrations
between 1100 and 1400 cm$^{-1}$ were determined by projection and the
results confirm mixing of these modes.\\

\noindent
In summary, the simulations confirm that the modes in the 1100 to 1400
cm$^{-1}$ frequency range in F-PhOH are strongly coupled. Assignment
of individual spectral features from comparing experiment with
simulations is not obvious. Consistent with experiment the force
field-based simulations for the high-frequency modes find a
high-frequency ($> 3600$ cm$^{-1}$) phenolic -OH stretch together with
broad features below 3000 cm$^{-1}$. These extended absorptions are
also found from QM MD simulations without, however, the high-frequency
-OH stretch.\\

\subsection{Comparison of the Spectroscopy for Hydrated PhOH and F-PhOH}
The experimental spectra for F-PhOH (black) and PhOH (green) in water
are reported in Figure \ref{fig:spec_solv}A. For the experimental
spectra between 1100 cm$^{-1}$ and 1400 cm$^{-1}$ pronounced
differences between the two compounds are found. Most prominently, the
single band with maximum at 1242 cm$^{-1}$ for PhOH, which is one of
the ``X-sensitive modes'' involving considerable motion of the phenol
ring and the CO group,\cite{evans:1960} is shifted to the red for
F-PhOH and split into at least two (at 1201 and 1222 cm$^{-1}$), but
possibly several more peaks, some of which overlap with the peak from
PhOH. Other features, such as the broader band with peak maximum at
1381 cm$^{-1}$ for PhOH are also shifted to the red (band maximum at
1368 cm$^{-1}$) for F-PhOH, see Figure \ref{fig:spec_solv}A.\\

\noindent
Consistent with experiment, the number of spectral features for F-PhOH
is larger than for PhOH in the 1200 to 1300 cm$^{-1}$ range. However,
none of the computed spectra display the pronounced double-peak
structure above 1200 cm$^{-1}$ for F-PhOH with the broad peak for PhOH
to the blue of it. The simulations using MTPs find a single absorption
between the low- and high-frequency absorption in F-PhOH, see Figure
\ref{fig:spec_solv}B (green). Considering the CO-power spectrum the
feature at 1224 cm$^{-1}$ involves the CO-stretch vibration together
with a band at 1283 cm$^{-1}$. The QM/MM simulations report a larger
number of spectroscopic features for PhOH (Figure
\ref{fig:spec_solv}D, green) than experiment does. In particular, the
single absorption at 1242 cm$^{-1}$ is not present but rather a broad
absorption extending from 1150 up to $\sim 1250$ cm$^{-1}$ is
found. Finally, simulations using the PhysNet energy function quite
well capture the absorption for PhOH at 1270 cm$^{-1}$ (Figure
\ref{fig:spec_solv}E, green) with an additional peak above 1300
cm$^{-1}$ not present in the experiment. Correcting the position 1270
cm$^{-1}$ by 0.94 as was done for the CH- and OH-stretch vibrations
shifts this band too far to the red compared with experiment, probably
because for coupled vibration the standard correction factor is
inappropriate. The absorption for F-PhOH is shifted to the red, in
agreement with experiment but does not exhibit the double peak
structure. Finally, it should be noted that the experiments are
carried out at solute concentrations that favour F-PhOH and PhOH
dimers and oligomers to be formed which can affect both, the position
of the absorption frequency and the line shape and make direct
comparison with simulations potentially difficult.\\

\noindent
For the high-frequency part (Figure \ref{fig:spec_solv}A right panel)
the spectra of solvated PhOH and F-PhOH follow each other closely
except for a pronounced absorption at 3596 cm$^{-1}$ in PhOH which
blue-shifts to 3643 cm$^{-1}$ upon fluorination. This suggests that
either electronic coupling between the CF- and OH-sites leads to a
slightly stronger OH-bond strength in F-PhOH compared with PhOH, or
that the hydration structure around -OH is affected by fluorination,
or a combination of the two. From normal mode calculations
(MP2/6-31G(d,p)) the OH-stretch vibrations are at 3833 cm$^{-1}$ and
3829 cm$^{-1}$ for PhOH and F-PhOH, which is an insignificant
difference and suggests that an electronic origin for the shift is
unlikely. For the MTP (3636 and 3588 cm$^{-1}$) and ML/MM (3794 and
3793 cm$^{-1}$) simulations the -OH stretch for PhOH and F-PhOH
reproduce the proximity of the two absorptions in solution, but in
reverse order compared with experiment. The QM/MM simulations for PhOH
(green trace Figure \ref{fig:spec_solv}D) report the OH-stretch at
3247 cm$^{-1}$ which is to the red of that for F-PhOH (3257 cm$^{-1}$,
orange trace), in agreement with experiment. However, the absorptions
are shifted by about 300 cm$^{-1}$ to the red relative to the
experimental line positions.\\

\subsection{Radial Distribution Functions and Solvent Distribution}
\begin{figure}[H]
\begin{center}
\includegraphics[width=0.9\textwidth]{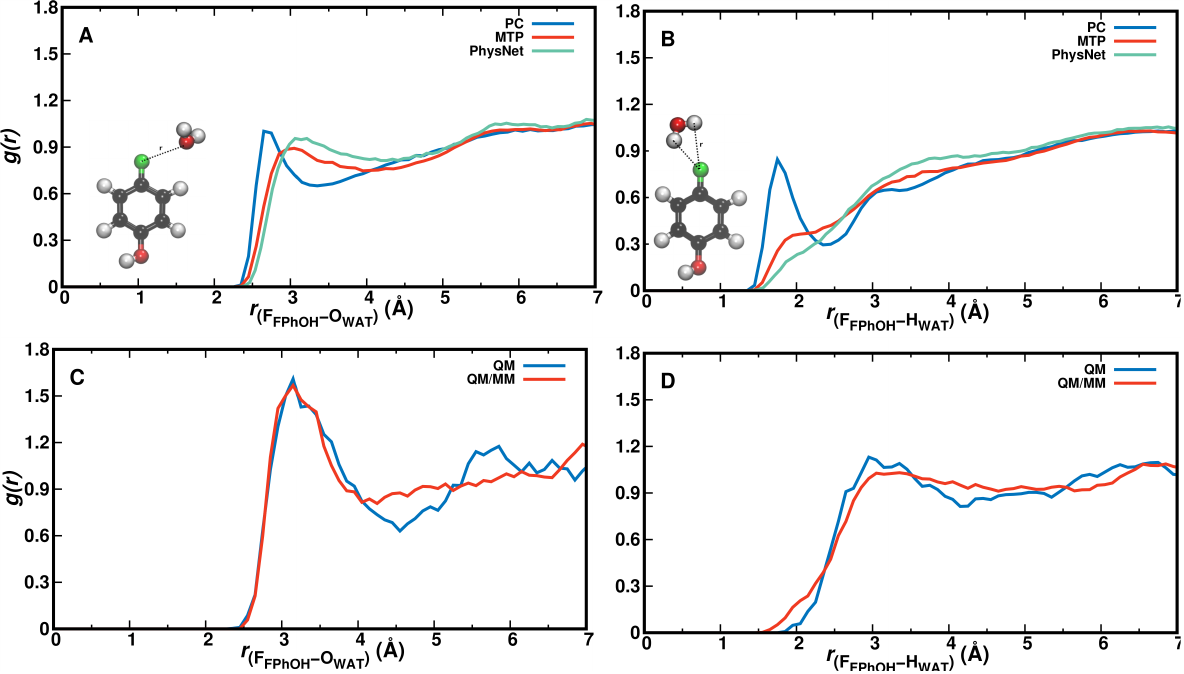}
\caption{The $g(r)$ for A) F–-O$_{\rm WAT}$ and B) F-–H$_{\rm WAT}$
  separations from PC (blue), MTP (red) and PhysNet (green)
  simulations of F-PhOH in H$_2$O. Using PCs both $g(r)$ are more
  structured whereas the radial distribution functions from MTP and
  PhysNet - which contains conformationally fluctuating charges - are
  rather similar to one another. Both, MTP and PhysNet, point to weak
  interaction between the fluorinated end and the environment. Panels
  C and D show similar The scaling along all $x-$axes is identical
  whereas that along the $y-$axis is not.}
\label{fig:grnr_fowat_fhwat}
\end{center}
\end{figure}

Radial distribution functions provide information about solvent-solute
interactions. For the F-O$_{\rm WAT}$ and F-H$_{\rm WAT}$ distances
they are reported in Figure \ref{fig:grnr_fowat_fhwat} for the PC,
MTP, and ML/MM MD simulations (panels A and B) and for the QM/MM and
QM simulations in panels C and D. For $g_{\rm F-O_W}$ the position of
the first maximum agrees quite favourably for all methods except for
simulations with PCs for which the maximum is shifted to shorter
separations and the first peak is unusually sharp. Qualitatively, the
four other methods find comparable shapes although the two QM
simulations have a more pronounced first minimum than MTP and ML/MM
simulations which may be related to a somewhat stronger interaction
between the fluorine and the water-oxygens or to the shorter sampling
time.\\

\noindent
The F-H$_{\rm WAT}$ pair correlation function $g_{\rm F-H_W}$ shows
even more pronounced differences between simulations with PCs compared
with all other models. The peak at 1.75 \AA\/ points towards a strong,
favourable interaction between solvent hydrogen atoms and the fluorine
atom which is not found in any of the other four methods. This can be
explained by the negative partial charge $q_{\rm F} = -0.29e$ on the
Fluorine atom in the PC model. The MTP and ML/MM simulations find
comparable distribution functions whereas the two QM-based simulations
differ from this in that the broad first maximum is peaked at around 3
\AA\/ with the full QM simulations and shows a more pronounced local
minimum. All distribution functions except that with PCs report the
first maximum at an F--O$_{\rm W}$ separation of $\sim 3$ \AA\/ which
points towards a largely hydrophobic behaviour of the CF site. This is
also consistent with notions from pharmaceutical chemistry in that a
CF group reduces ligand solvation and increases its
hydrophobicity.\cite{Rachel:2020} The number of water molecules within
distance $r$ is reported in Figure S4 and shows that for
small F--water separations ($r \leq 4$ \AA\/) the occupation from
simulations with MTP and PhysNet is similar but clearly below that of
QM/MM and QM simulations (which are identical) whereas for larger
separations ($r \sim 5$ \AA\/) that from QM simulations approaches the
MTP and PhysNet simulations.\\

\begin{figure}[H]
\begin{center}
\includegraphics[width=0.9\textwidth]{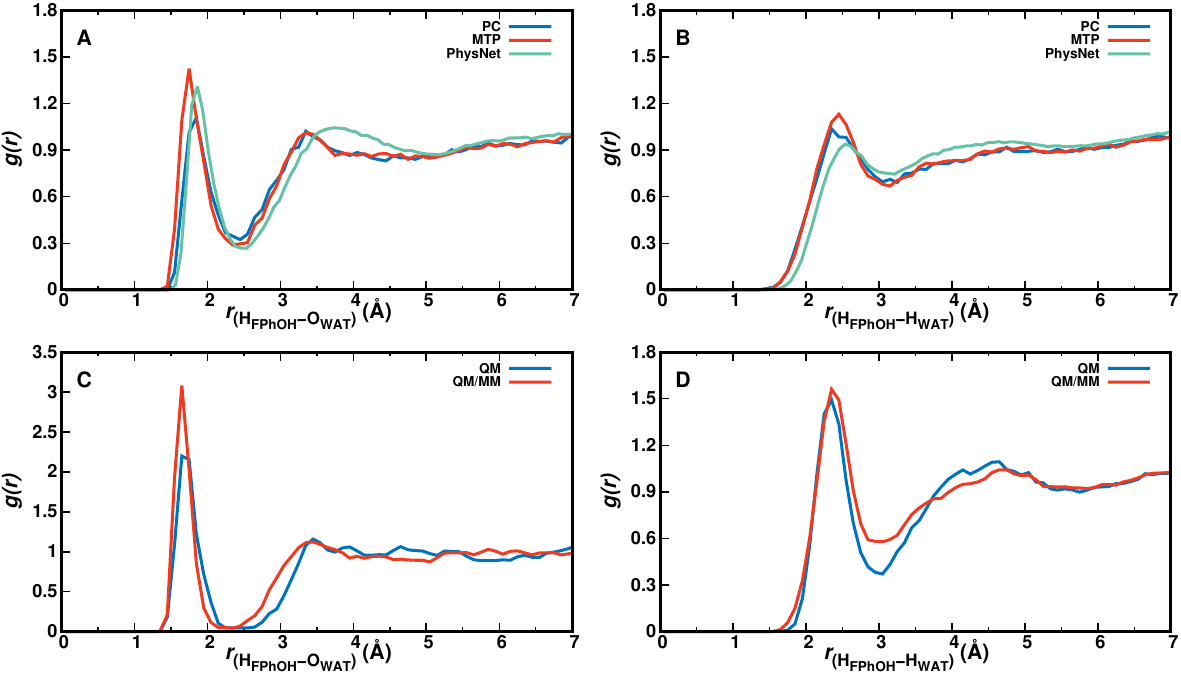}
\caption{The $g(r)$ for A) H$_{\rm FPhOH/PhOH}$–-O$_{\rm WAT}$ and B)
  H$_{\rm FPhOH/PhOH}$-–H$_{\rm WAT}$ distances between the H-atom of
  the solute OH-group as obtained from PC (blue), MTP (red) and
  PhysNet (green) simulations for F-PhOH in H$_2$O. Panels C and D
  from QM and QM/MM simulations. The scaling along all $x-$axes is
  identical whereas that along the $y-$axis it is not.}
\label{fig:grnr_howat_hhwat}
  \end{center}
\end{figure}

\noindent
For analyzing hydration around the hydroxyl group the H$_{\rm
  OH}$--O$_{\rm WAT}$ and H$_{\rm OH}$--H$_{\rm WAT}$ pair correlation
functions were considered, see Figure \ref{fig:grnr_howat_hhwat}. From
simulations using the PC, MTP, and PhysNet models $g_{\rm H_{OH}-O_W}$
characterizing the hydrogen bond between OH and water-oxygen atoms is
similar up to and including the first minimum, see Figure
\ref{fig:grnr_howat_hhwat}A. The second maximum shifts to somewhat
larger separations for the PhysNet simulations. Similar observations
are made for $g_{\rm H_{OH}-H_W}$ in panel B. The hydrogen-bond pair
distribution function with a first maximum at $\sim 1.6$ \AA\/
suggests that solvent water molecules are quite strongly bound to the
-OH group and that the solvent water at the --OH group exchanges.\\

\noindent
For $g_{\rm H_{OH}-O_W}$ from the QM and QM/MM simulations (blue and
red traces in Figure \ref{fig:grnr_howat_hhwat}C) the position of the
first maximum is almost identical whereas the height of the first peak
differs. Both pair correlation functions have a first minimum around
2.5 \AA\/ with an amplitude close to 0 which suggests that during the
25 ps simulation one solvent water molecule is strongly bound to the
OH-group of F-PhOH and does not exchange with the surrounding
solvent. As a consequence the ``free phenolic -OH stretch''
(spectroscopic feature around and above 3600 cm$^{-1}$, Figure
\ref{fig:spec_solv}) is absent in the QM MD simulations, in contrast
with what was found from the experiments and from the PC, MTP, and
ML/MM MD simulations. On the other hand, the radial distribution
function between the water-oxygen atoms and the two carbon atoms
flanking the COH group in F-PhOH from a 5 ns MTP simulation (Figure
S5) demonstrates that the solvent distribution
dynamics is exhaustively sampled on the 5 ns time scale.\\

\noindent
For the phenolic oxygen as an H-bond acceptor the O--H$_{\rm WAT}$ and
O--O$_{\rm WAT}$ pair correlation functions for F-PhOH are reported in
Figure S6. Panels A and B provide a direct
comparison of the three force field-based simulations whereas panels C
and D are those from the QM and QM/MM simulations, respectively. The
MTP simulations (red) find strongest localization of the water
followed by PC (blue) and PhysNet (green) simulations. For the QM
simulations the first minimum for $g_{\rm O-O_W}$ is deeper than for
the QM-MM simulations whereas the position and height of the first
maximum are comparable. The radial distribution functions from
simulations with MTP are closest to those from the QM and QM/MM
simulations, respectively.\\

\begin{figure}[H]
\begin{center}
\includegraphics[width=0.9\textwidth]{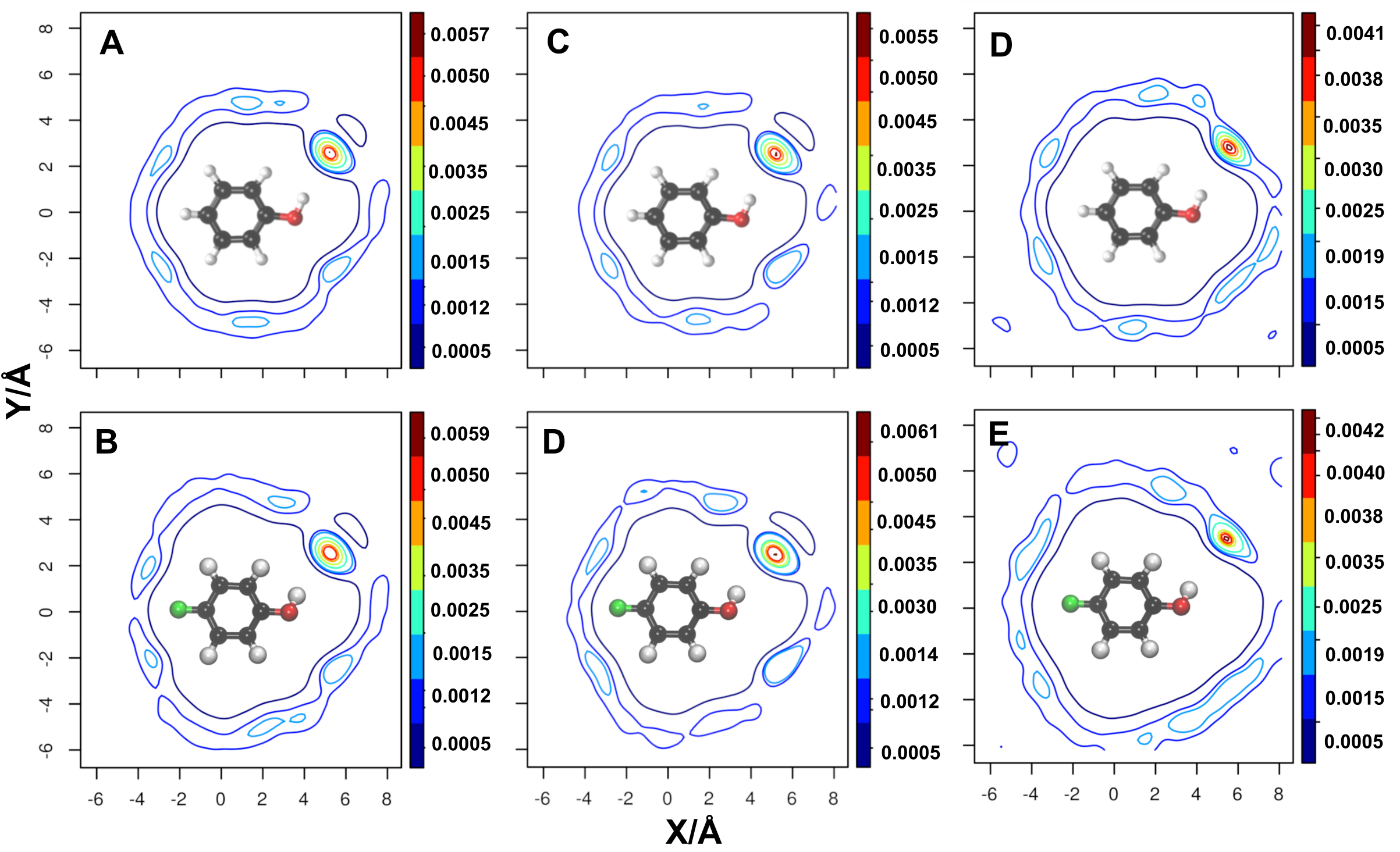}
\caption{2-dimensional solvent distribution from reorienting snapshots
  with respect to all heavy atoms of the solute, including the
  phenolic oxygen atom. Solvent distributions around PhOH and F-PhOH
  for PC (panel A and B), MTP (panel C and D) and from PhysNet (panel
  E and F) model. The iso-contour values are shown in each panel. For
  the solvent distribution upon reorienting with respect to all heavy
  atoms excluding the phenolic oxygen atom, see Figure
  S7 which manifestly shows symmetric water
  distribution according to the underlying dynamically averaged
  spatial symmetry of the solute due to rotation of the OH group
  around the -COH axis, see Figure S8.}
\label{fig:solv_dist}
\end{center}
\end{figure}

\noindent
Two-dimensional solvent distribution functions were generated from the
positions of the water-oxygen atoms around F-PhOH. For that, the
structures of the 5000 snapshots were oriented with C1 in the origin,
the C1--C4 bond along the $x$-axis the [C1,C4,H] atoms in the
$xy-$plane. A 2-dimensional histogram of the water positions was
generated and then refined from kernel density estimation using
Rstudio.\cite{rstudio} The distribution of the solvent water around
the CF-part of F-PhOH is comparatively flat for all simulations with
PC, MTP and PhysNet when contrasted with the COH-moiety of the solute,
see Figure \ref{fig:solv_dist}. It is also found that for PhOH (top
line) the solvent distributions resemble those for F-PhOH (bottom
line). The asymmetry in the solvent distribution around the -COH group
is due to the reference atoms chosen for the superposition of all
structures. If the phenol-oxygen atom is excluded in reorienting the
structures the 2d distribution becomes manifestly symmetric (Figure S7)
 and also clarifies that the -OH group of the
solute rotates on the time scale of the simulations, see Figure
S8. Specifically, rotation of the -COH group is an
activated process and occurs on the $\sim 100$ ps time scale from MTP
and ML/MM simulations.\\

\noindent
The weak interaction between solvent and solute around the CF-site of
F-PhOH can also be gleaned from the behaviour of the frequency
fluctuation correlation functions (FFCFs). For this, the FFCFs were
determined from the frequency trajectories $\omega_{i}(t)$ for the
five modes in the CF-stretch region, i.e. $\nu_1$ to $\nu_5$, between
1100 cm$^{-1}$ and 1400 cm$^{-1}$ and for the OH-stretch vibration,
see Figures S9 and S10. The FFCFs contain
information about the coupling between a particular mode and the
environmental dynamics. The fitting parameters for $\nu_1$ to $\nu_5$
in Table S5 show that the fast correlation is generally
$\tau_1 \sim 0.1$ ps whereas the longer time scale ranges from $\tau_2
=0.28$ ps to $\tau_2 =0.83$ ps. Such short correlation times point
towards weak solvent/solute interactions close to the CF-site. The
static components $\Delta_0$ are very small, too, which also imply
rapid, unspecific dynamics around the CF-group. For the OH-stretch
vibration, the fast correlation time is $\tau_1 \sim 0.1$ ps while the
longer one is $\tau_1 \sim 0.5$ ps which are comparatively
short. Compared with the FFCFs for vibrations involving the CF-stretch
the $t=0$ amplitude and the static offset $\Delta_0$ is larger by one
to two orders of magnitude which points towards somewhat slower
dynamics around the OH bond of F-PhOH.\\

\section{Discussion and Conclusion}
The present work discusses the infrared spectroscopy of and solvent
distribution around F-PhOH in water from experiments and a range of
computational approaches including full QM, QM/MM, PC-based, MTP-based
and ML/MM simulations. The simulations show that the modes in the
range between 1200 and 1300 cm$^{-1}$ are heavily mixed which
complicates the assignment of the spectroscopic features. Accounting
for the known overestimation of frequencies from MP2/6-31G(d,p)
calculations the ML/MM simulations best capture the experimentally
observed patterns in this frequency range. The solute-solvent pair
distribution functions together with the FFCFs indicate that
interaction between the fluorinated position and the environment is
weak as has also been found earlier for
fluoro-acetonitrile.\cite{MM.facn:2015}\\

\noindent
For the phenol-OH-stretch vibration both, ``gas phase'' (around 3600
cm$^{-1}$) and ``water-phenol hydrogen bonded'' signatures (below 3000
cm$^{-1}$) are found from experiments and the simulations. A regular
pattern with a spacing of $\sim 50$ cm$^{-1}$ is reminiscent of recent
SFG spectra of PhOH at the air/water interface and earlier experiments
on liquid and solid pure PhOH\cite{evans:1960} and is assigned to
water-phenol hydrogen bonded motions. For the high frequency
(OH-stretch) modes early experiments for PhOH vapor, in apolar solvent
(CCl$_4$) and as a liquid reported band positions at $\sim 3650$
cm$^{-1}$, $\sim 3600$ cm$^{-1}$, and at 3500 cm$^{-1}$,
respectively.\cite{evans:1960} In CCl$_4$ solution and the pure liquid
an additional band is at 3350 cm$^{-1}$.\cite{evans:1960} The present
work finds a sharp peak at 3596 cm$^{-1}$ which is assigned to the
``free'' phenol-OH stretch in a non-hydrogen bonded environment. This
is supported by the observation that the peak is only marginally
shifted from the PhOH OH-stretch in vapor and the fact that the
spectroscopic feature is sharp and therefore can not be due to
water.\\

\noindent
More recently, infrared spectra were recorded for PhOH complexed with
variable numbers of water molecules in the gas
phase\cite{hamashima:2011,shimamori:2015}, in
matrices,\cite{banerjee:2017} and for PhOH at the air/water interface
using vibrational sum frequency generation (SFG).\cite{tahara:2018}
The cluster studies all report the phenolic-OH stretch vibration at
frequencies above 3000 cm$^{-1}$ whereas the experiment at the
air/water interface assigns a very broad signature in the SFG signal
extending from 2550 cm$^{-1}$ to 3500 cm$^{-1}$ to the OH-stretch
mode.\cite{tahara:2018} This finding is consistent with the present
experiments which report a broad absorption for both, F-PhOH and PhOH
extending down to $\sim 2700$ cm$^{-1}$ which is also assigned to the
phenol-OH stretch for water-coordinated -COH, see Figures
\ref{fig:spec_solv} and S2.\\

\noindent
Hydration around the -CF group as characterized by 1d- and 2d-solvent
distribution functions does not feature any pronounced hydrogen
bonding unless a conventional PC model is used which, however,
exaggerates the directed interaction along the CF bond. On the other
hand, solvent water molecules form intermittent H-bonds with the -OH
group of F-PhOH which leads to the broad spectroscopic response below
3000 cm$^{-1}$ observed experimentally and reproduced by several of
the computational models.\\

\noindent
From the perspective of halogenic modifications of organic frameworks
used for drug design the present work suggests that the local
hydration of -CH and -CF groups is comparable (see Figures
\ref{fig:grnr_fowat_fhwat} and S5), supporting the
notion that ``F behaves like a large H-atom''.\cite{Rachel:2020}
Although the C-F bond has three isolated electron pairs, it has weaker
electrostatic interactions compared to O due to the small size and
high electronegativity, which compromises its hydrogen bonding ability
and can thus better be described as a weakly polar interaction rather
than a hydrogen bond. Such insights are important guidelines for
rational drug discovery as they provide a basis for directed
modification and evolution of ligands with specific interactions in
protein binding sites.\\

\begin{suppinfo}
The supporting information provides tables for the force field
parametrizations (Tables S1 to S3),
frequency maxima for frequency distributions (Table S4), and
 parameters for the FFCFs (Table S5) together with Figures S1 to
S10 the report quality of the PhysNet model, additional
solvent distribution functions, dihedral time series and the FFCFs for
the CF- and OH-stretch frequencies.
\end{suppinfo}

\section{Acknowledgments}
This work was supported by the Swiss National Science Foundation, NCCR
MUST (to MM, PH and UR).\\

\bibliography{refs}
\end{document}


\begin{table}[H]
\begin{tabular}{|c|c||c|c|}
 \hline\hline
 \hline\hline
 Atom (PhOH) & Charge ($e$) & Atom (F-PhOH)& charge ($e$) \\
 \hline 
   C1 &  -0.095 & C1 & 0.161 \\
     \hline 
   C2 &  -0.074 & C2 & -0.060 \\
     \hline 
   C3 &  -0.079 & C3 &  -0.061 \\
     \hline 
   C4 &  0.075 & C4 &  0.064 \\
  \hline 
   C5 &  -0.079 & C5 & -0.061 \\
    \hline 
   C6 &  -0.074 & C6 &  -0.060 \\
    \hline 
   H7 &   0.086 & H7 &  0.113 \\
    \hline 
   H8 & 0.102 & H8 &  0.105 \\
    \hline 
   H9 & 0.102 & H9 &  0.105 \\
    \hline 
  H10 &  0.086 & H10 &  0.113 \\
   \hline 
  O11 &  -0.392 & O11 &  -0.389 \\
   \hline 
  H12 &  0.259 & H12 & 0.260 \\
   \hline 
  H13 &  0.082 & F13 &  -0.291 \\
 \hline
 \hline
\end{tabular}
\caption{Molecular monopoles calculated using a fitting environment
  with GDMA algorithm for PhOH and F-PhOH.\cite{mm.mtp2:2016}}
\label{sitab:chg}
\end{table}

\begin{table}[H]
\begin{tabular}{|c|c|c|c|}
 \hline\hline
 Atom & Q10 ($e$)&Q11c ($e$)&Q11s ($e$)\\
  \hline\hline
  \multicolumn{4}{|c|}{PhOH} \\
 \hline \hline
C1 & -0.015 &0.0 &0.033\\
\hline
C2 &-0.016&  - & 0.030\\
\hline
C3 &-0.0002 & -  & 0.054\\
\hline
C4 &0.028& 0.0& 0.089\\
\hline
C5 &-0.0002& - & 0.054\\
 \hline
C6 &-0.016& - & 0.030\\
\hline
O11 & 0.0& 0.099& -0.070  \\
\hline 
H13 &  0.0 & 0.0 &  0.0 \\
\hline \hline 
\multicolumn{4}{|c|}{F-PhOH} \\
\hline\hline
C1 & -0.011 &0.0&0.180\\
\hline
C2& -0.022&  -  & 0.073\\
\hline
C3&-0.025&  - & 0.017\\
\hline
C4 &0.015& 0.0& 0.122\\
\hline
C5 &-0.025& - & 0.017\\
 \hline
C6&-0.022& - & 0.073\\
\hline
O11&0.0 &0.121& -0.140\\
\hline 
F13& 0.147& 0.0& 0.008\\
 \hline
\end{tabular}
\caption{Atomic dipoles for PhOH and F-PhOH from fitting to the
  molecular electrostatic potential.\cite{mm.mtp2:2016} Q$xx$ are the
  spherical MTP coefficients expressed in the local axis system.}
\label{sitab:dip}
\end{table}

\begin{table}[H]
\begin{tabular}{|c|c|c|c|c|c|}
 \hline\hline
 Atom & Q20 ($e$)&Q21c ($e$)&Q21s ($e$)&Q22c($e$)&Q22s($e$) \\
  \hline\hline
 \multicolumn{6}{|c|}{PhOH} \\
 \hline\hline
C1 & -0.029 &0.0 &0.001&-0.005& 0.0\\
\hline 
C2 &-0.026& - & -0.0003& -0.0013& - \\
\hline 
C3 &-0.012& - & 0.002& 0.0039& - \\
\hline 
C4 &8.04$\times$10$^{-5}$ & 0.0& 0.013& 0.0015& 0.0\\
\hline 
C5 &-0.012& - & 0.0029 &0.0039 & - \\
\hline 
C6 &-0.0260& -  & -0.0003& -0.0013 & - \\
\hline 
O11 & -0.006& 0.0& 0.0& 0.015& -0.0278\\
\hline 
H13 & 0.0 & 0.0 &  0.0 &0.0 &0.0 \\
\hline\hline
 \multicolumn{6}{|c|}{F-PhOH} \\
\hline\hline
C1 &-0.024& 0.0& 0.006& -0.007& 0.0\\
\hline 
C2 &-0.042& - & 0.003& 0.0005& - \\
\hline 
C3&-0.029& -  & 0.002& 0.009& - \\
\hline 
C4&0.005& 0.0& 0.022& -0.009& 0.0\\
\hline 
C5&-0.029& - &0.002& 0.009& - \\
\hline 
C6&-0.042& - & 0.003& 0.0005 & - \\
\hline 
O11&-0.034& 0.0 & 0.0& 0.029& -0.0798\\
\hline 
F13 &-0.034& 0.0& 0.0016& 0.009& 0.0\\
\hline
\end{tabular}
\caption{Atomic quadrupoles for PhOH and F-PhOH from fitting to the
  molecular electrostatic potential.\cite{mm.mtp2:2016} Q$xx$ are the
  spherical MTP coefficients expressed in the local axis system.}
\label{sitab:quad}
\end{table}

\begin{table}[H]
\footnotesize
\centering
\caption{The frequency maxima obtained from frequency
    distribution, power and IR spectrum using PC/MTP model in the gas
    phase and solvent for five different modes in the frequency range
    of 1100 to 1400 cm$^{-1}$ .}  \centering
\begin{tabular}{|r|cc|cc|cc|cc|}
\hline
\multicolumn{9}{|c|}{MD} \\
\hline
   \multirow{2}{*}{Mode} &
      \multicolumn{2}{c}{$P(\omega)$} &
      \multicolumn{2}{c}{PS(CF)} &
      \multicolumn{2}{c|}{PS(CF)} &
      \multicolumn{2}{c|}{IR(CF)} \\
& MTP$_{\rm gas}$& MTP$_{\rm H_{2}O}$ & MTP$_{\rm gas}$& MTP$_{\rm H_{2}O}$ & PC$_{\rm gas}$& PC$_{\rm H_{2}O}$ & PC$_{\rm H_{2}O}$& MTP$_{\rm H_{2}O}$\\
    \hline
    \textbf{$\nu_{1}$} & 1140 & 1149 & 1141 & 1158 & 1140 & 1163 & 1148& 1151\\
    \hline
    \textbf{$\nu_{2}$} & 1170 & 1172 & 1170 & 1185 & 1171 & 1184 & 1172 & 1174\\
    \hline
    \textbf{$\nu_{3}$} & 1225 & 1236/1251 & 1227 & 1248 & 1226 & 1265 & 1243 & 1249\\
    \hline
    \textbf{$\nu_{4}$} & 1286 & 1294 & 1285 & 1305 & 1286 & 1310 & 1294 &1296\\
    \hline
    \textbf{$\nu_{5}$} & 1324 & 1330 & 1324 & 1345 & 1324 & 1343 & 1331 &1333\\
    \hline
\end{tabular}
\label{sitab:sshift}
\end{table}

\begin{table}[H]
\centering
\caption{Parameters obtained from fitting the FFCF to
  Eq. 3 based on frequencies from INM using 5 ns MTP
  ($10^6$ snapshots) simulation of F-PhOH in H$_{2}$O. Average
  frequency $\langle\omega\rangle$ of the asymmetric stretch in
  cm$^{-1}$, the amplitudes $a_1$ and $a_2$ in ps$^{-2}$, the decay
  times $\tau_1$ and $\tau_2$ in ps, and the offset $\Delta_0$ in
  ps$^{-2}$. The FFCF for the OH-stretch is reported in Figure S10.}

\begin{tabular}{l|c|cccc|c}
\hline\hline
Mode & $\langle\omega\rangle$ & $a_{1}$ &$\tau_{1}$ &$a_{2}$ &$\tau_{2}$ & $\Delta_0$ \\
\hline
\textbf{$\nu_1$} &1149.71  & 0.36  & 0.08   &   0.06   &  0.45  &    0.0005  \\
\textbf{$\nu_2$} &1173.00 & 0.29  &   0.05   &   0.04  &  0.38  &    0.0003   \\
\textbf{$\nu_3$} &1248.40  & 2.31  &  0.10   &   0.32  &   0.60  &    0.0095  \\
\textbf{$\nu_4$} &1295.18 & 0.68  &  0.08   &   0.14  &   0.28 &    0.0029   \\
\textbf{$\nu_5$} &1332.04 & 1.01  &  0.09   &   0.06  &   0.83  &    0.0055 \\
OH                         &3500.22 &32.11 & 0.10   &  17.72   & 0.50  &   0.19\\
\hline
\end{tabular}
\label{sitab:ffcffit}
\end{table}

\begin{figure}[H]
\begin{center}
\includegraphics[width=0.5\textwidth]{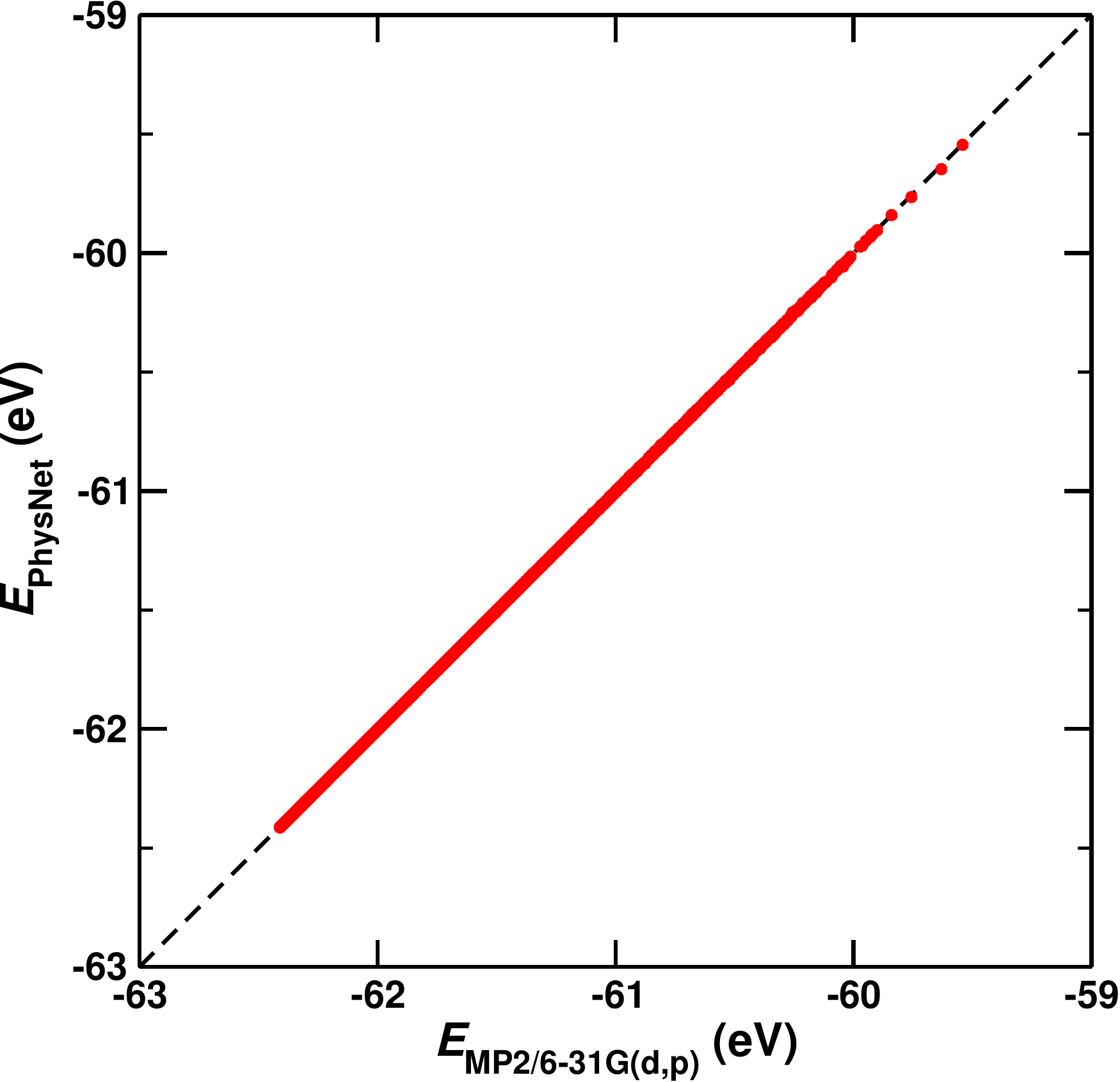}
\caption{Correlation between the ab initio and PhysNet energies for a
  set of 3700 randomly selected points averaged over 980/982
  trajectories with $R^2 = 0.9999$ and an root mean square error of
  0.0037 eV.}
\label{sifig:corrpes}
\end{center}
\end{figure}

\begin{figure}[H]
\begin{center}
\includegraphics[width=0.5\textwidth]{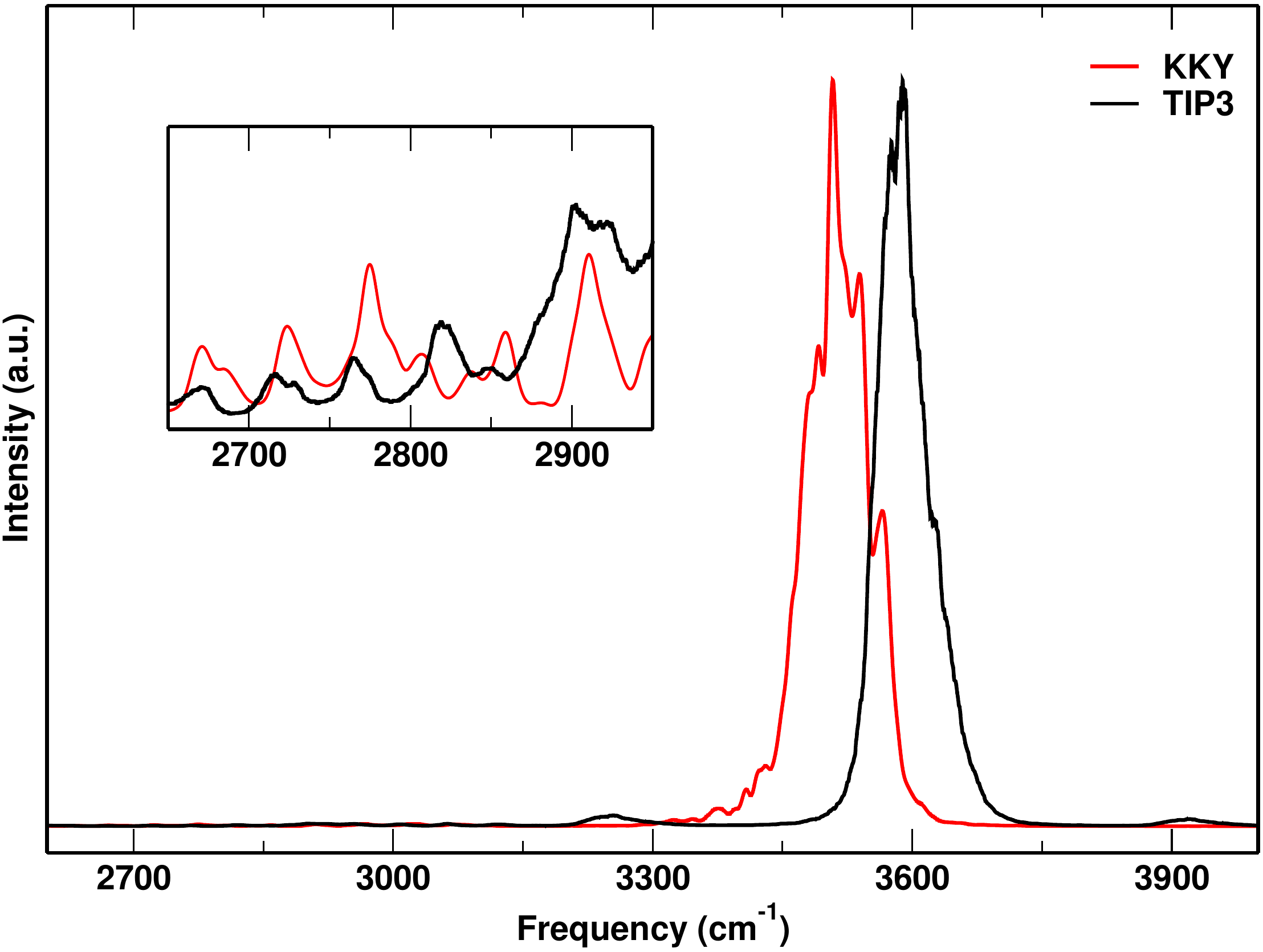}
\caption{The power spectrum for the phenol-OH stretching vibration
  from a 5 ns simulation of F-PhOH with flexible (red) KKY
  water\cite{Burnham97p6192,MM.ice:2008,kky_orig} compared to TIP3
  (black) model. The red shift from the simulations with KKY is --78
  cm$^{-1}$.}
\label{sifig:kky}
\end{center}
\end{figure}

\begin{figure}[H]
\begin{center}
\includegraphics[width=0.4\textwidth, angle=-90]{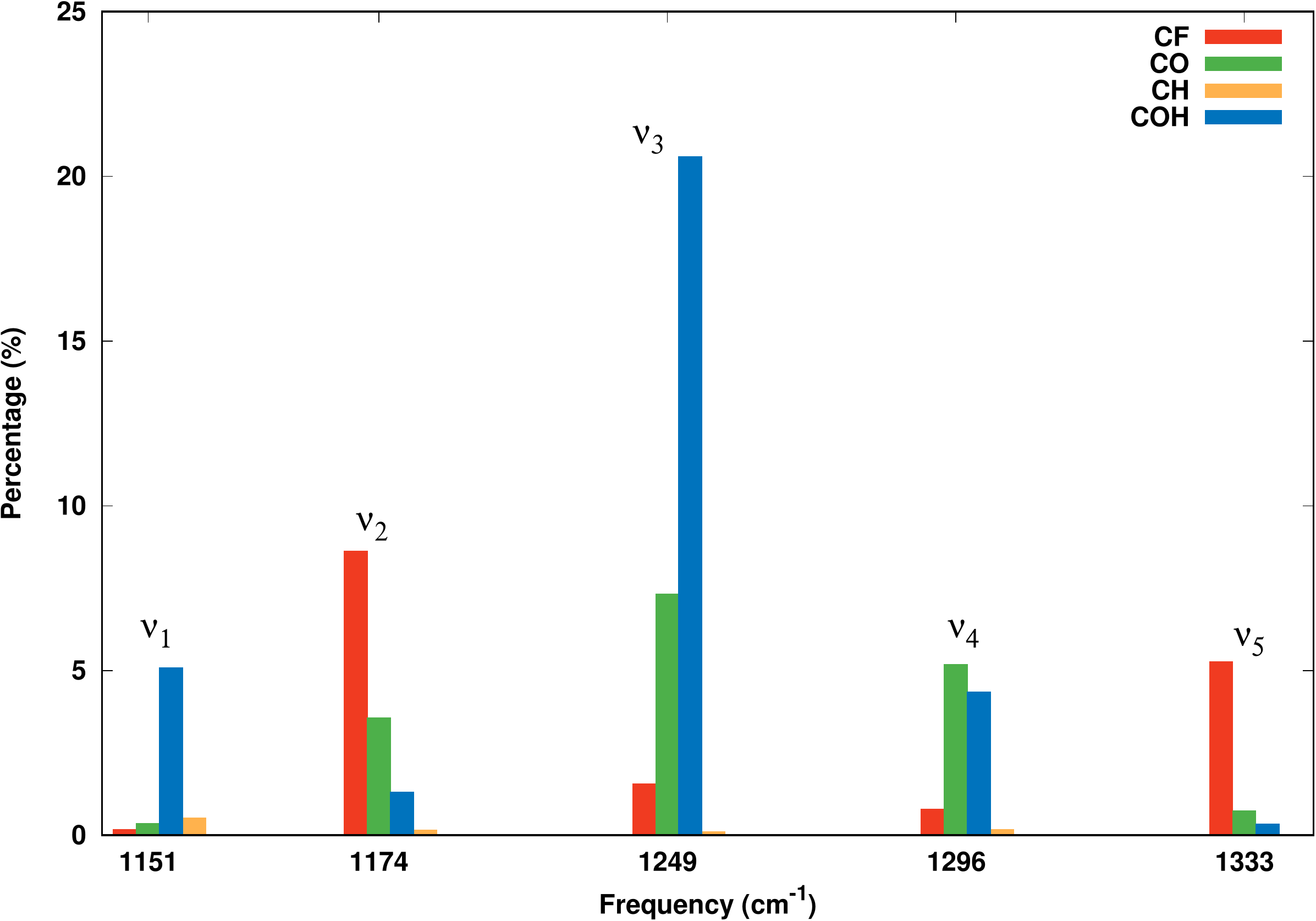}
\caption{Participation ratio of the CF (red), CO (green), and CH
  (orange) stretching and the COH bending (blue) motions to the 5
  modes between 1140 and 1350 cm$^{-1}$ by using the ``project''
  facility in CHARMM for $10^5$ snapshots from the MTP simulation of
  F-PhOH in H$_2$O. The remaining contributions are from low frequency
  modes.}
\label{sifig:proj}
\end{center}
\end{figure}

\begin{figure}[H]
\begin{center}
\includegraphics[width=0.4\textwidth,angle=-90]{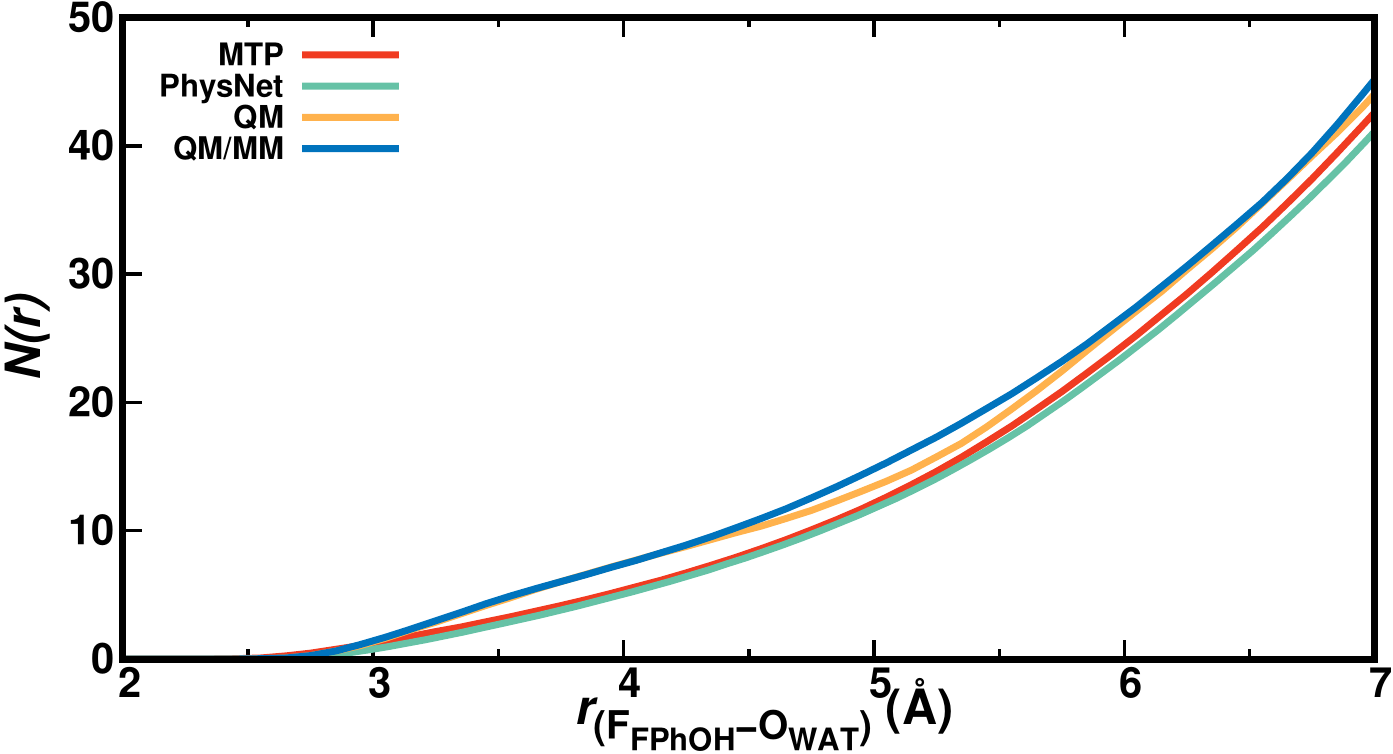}
\caption{The total number of water molecules within distance $r$ of
  the fluorine atom for the MTP (red), PhysNet (green), QM/MM (orange)
  and QM (blue) simulations.}
\label{sifig:nwat}
\end{center}
\end{figure}

\begin{figure}[H]
\begin{center}
\includegraphics[width=0.7\textwidth]{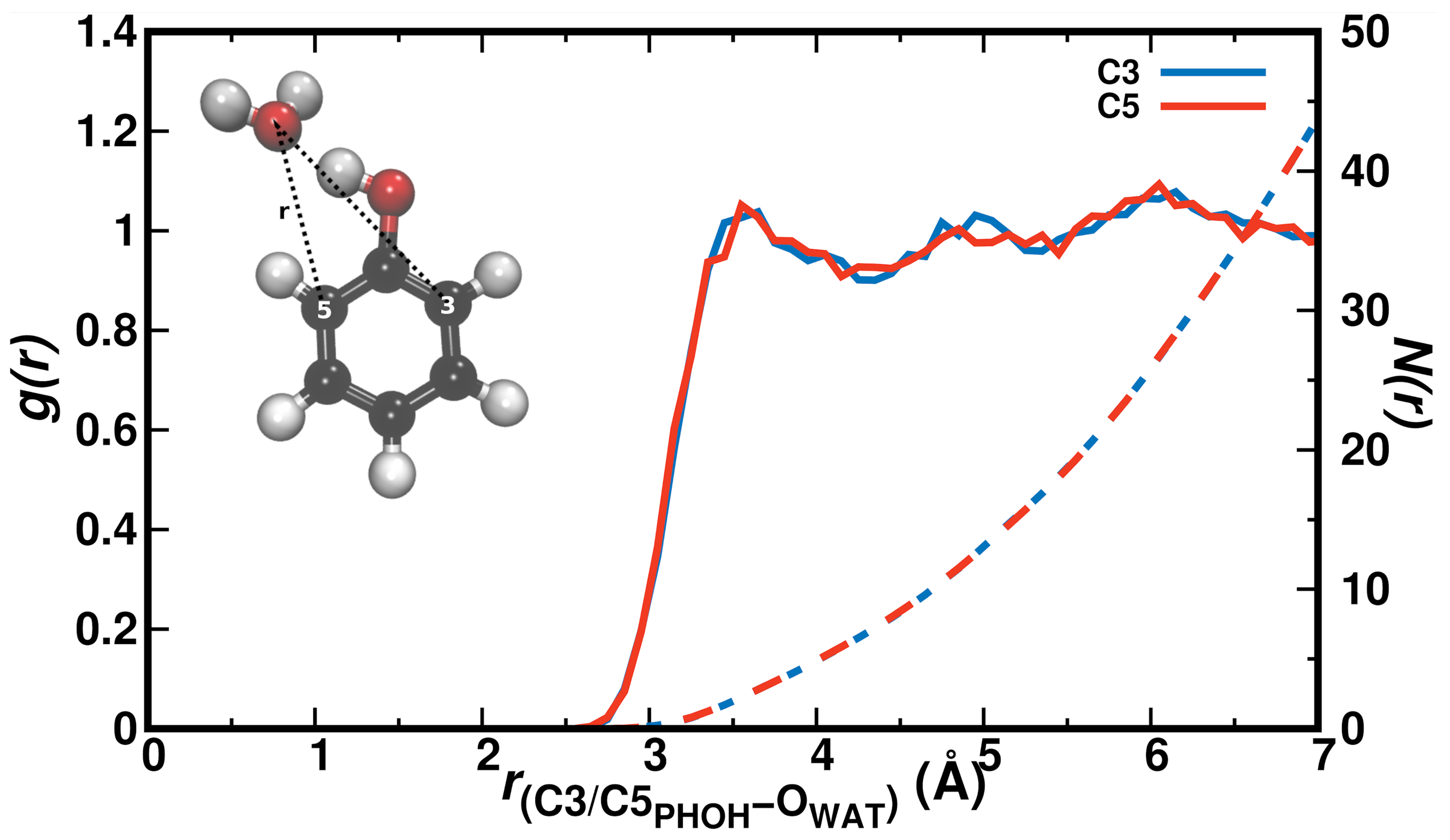}
\caption{The radial distribution function between the water-oxygen
  atoms and the two carbon atoms flanking the COH group in PhOH from a
  5 ns simulation with MTP. It is demonstrated that the solvent
  distribution is converged. Comparison with the red trace in Figure
  4A indicates that hydration around the CF
  and CH groups is similar with the first maximum at a similar
  C--O$_{\rm W}$ separation.}
\label{sifig:grcwat}
\end{center}
\end{figure}

\begin{figure}[H]
\begin{center}
\includegraphics[width=0.9\textwidth]{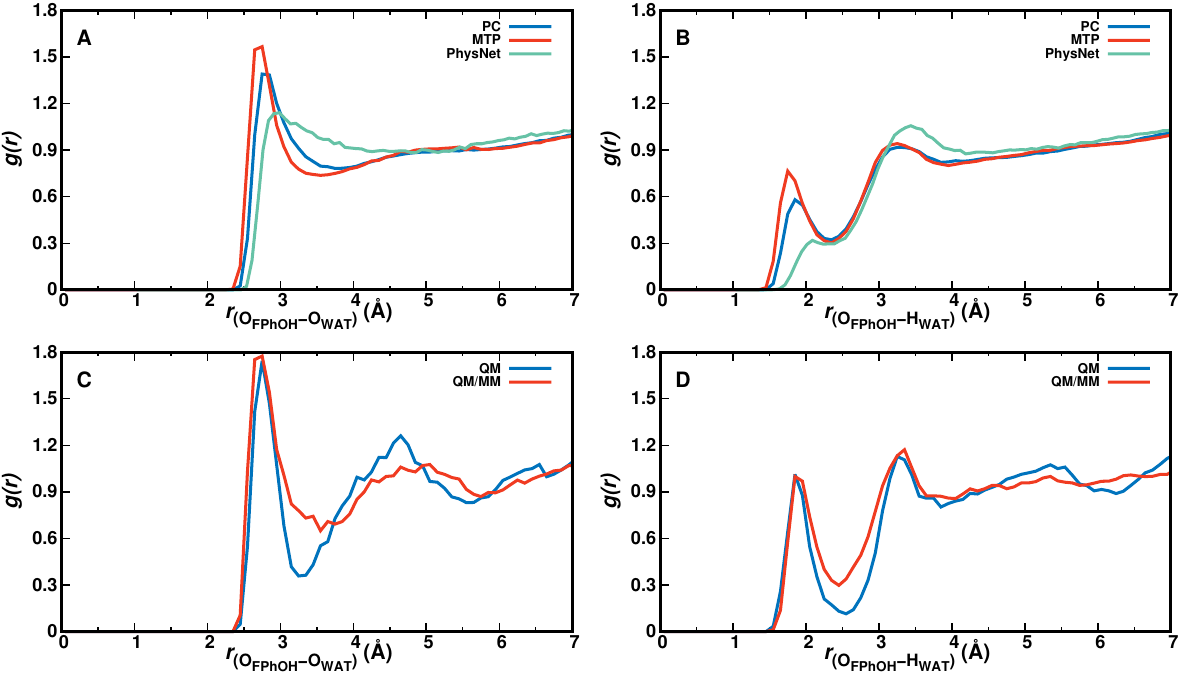}
\caption{The $g(r)$ for A) O$_{\rm FPhOH}$–-O$_{\rm WAT}$ and B)
  O$_{\rm FPhOH}$-–H$_{\rm WAT}$ distances as obtained from PC (blue),
  MTP (red) and PhysNet (green) simulations of F-PhOH in H$_2$O. Panel
  C and D from QM and QM/MM simulations.}
\label{sifig:grnr_oowat_ohwat}
  \end{center}
\end{figure}

\begin{figure}[H]
\begin{center}
  \includegraphics[width=0.5\textwidth]{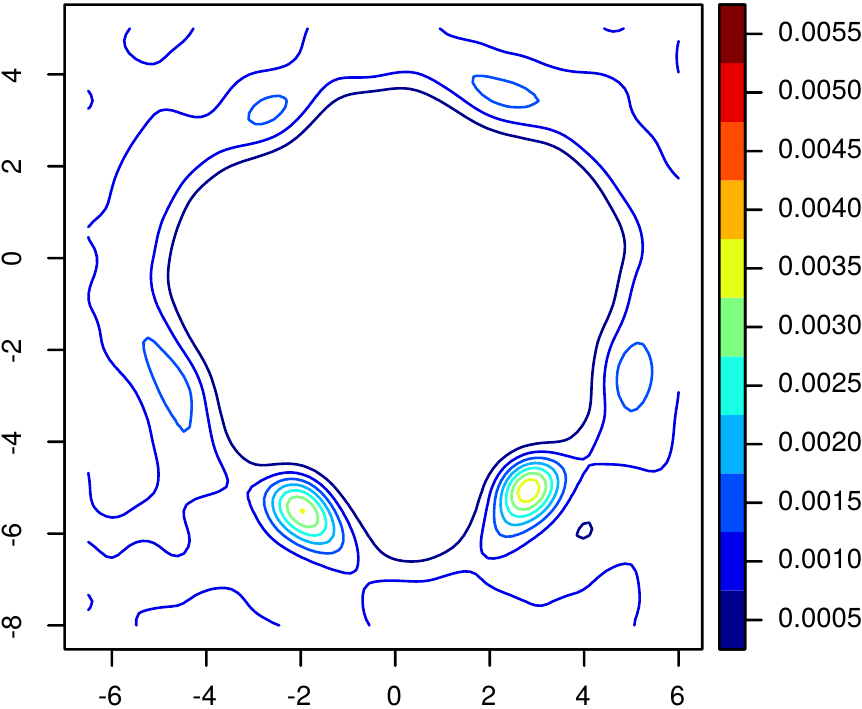}
  \caption{2d-distribution from MTP simulations with TIP3P water. The
    structures are reoriented with reference to the solute heavy
    atoms, but excluding the phenolic-oxygen atom. This differs from
    the selection used in Figure 6 and allows to
    visualize the symmetric distribution of the solvent water molecules
    around the -COH group.}
\label{sifig:2ddist}
  \end{center}
\end{figure}

\begin{figure}[H]
\begin{center}
\includegraphics[width=0.7\textwidth]{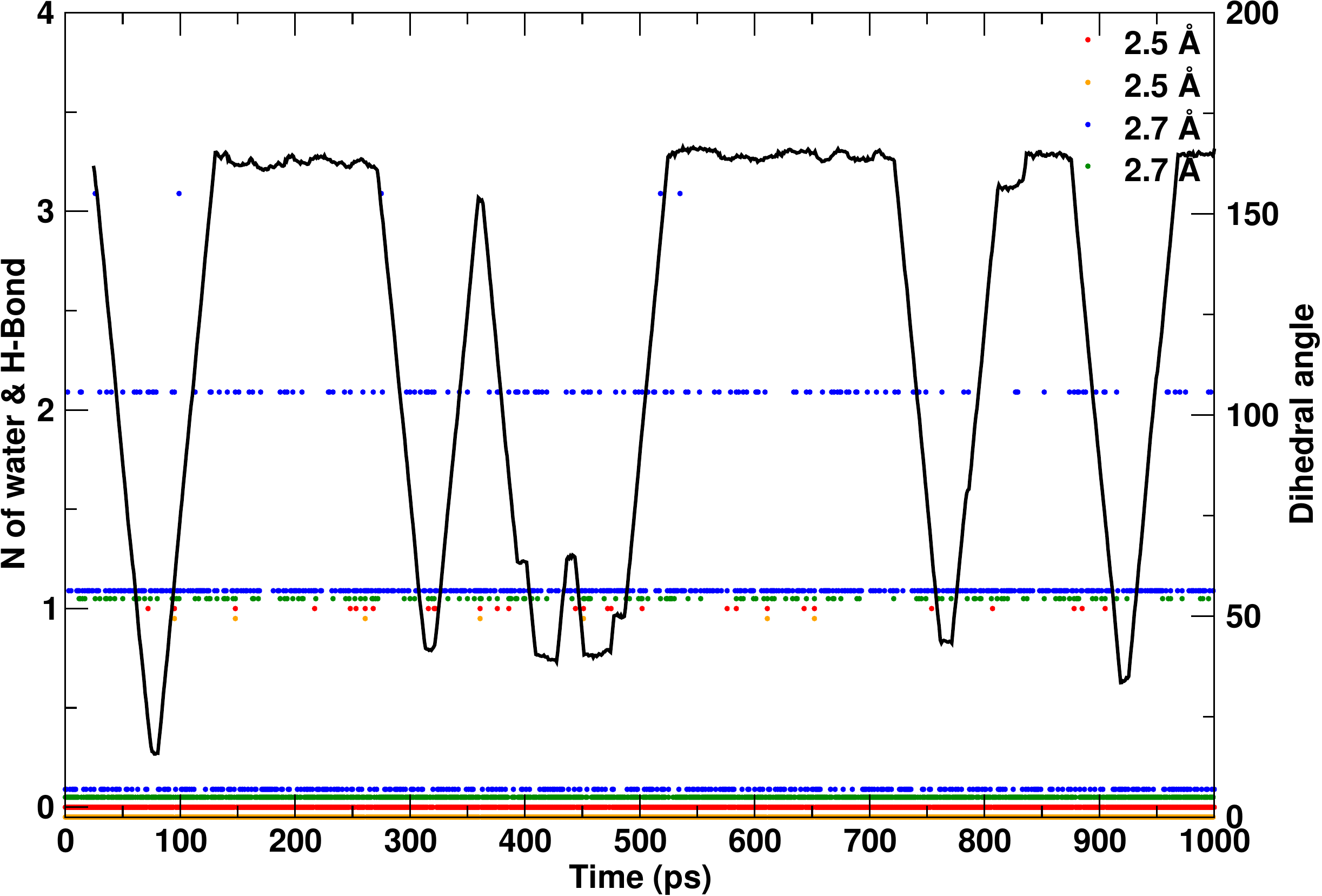}
\caption{Dihedral rotation of the CCOH angle together with the water
  occupation of the -OH group (circles) and number of H-bonds
  (crosses) for O$_{\rm COH}$--O$_{\rm W}$ separations of a) 2.5 \AA\/
  (red \& orange) and b) 2.7 \AA\/ (blue \& green). With the shorter
  cutoff there is no water molecule hydrogen bonded to the rotatable
  COH group for most of the simulation time and occasionally one water
  is sufficiently close to form an H-bond. For the longer cutoff water
  presence is frequent and H-bounding (green crosses) is more
  prevalent. The alteration between H-bonded and non-H-bonded -COH
  group explains the spectroscopy seen in Figure S2. A
  direct correlation between rotation of the CCOH angle and water
  occupation can not be established convincingly although it is likely
  that rotation of the COH group requires water to detach from it. The
  dihedral time series is locally averaged over a time window of 50
  ps. Simulations with ML/MM MD yield a comparable time series with
  transitions on the $\sim 100$ ps time scale.}
\label{sifig:dihedral}
\end{center}
\end{figure}

\begin{figure}[H]
\begin{center}
\includegraphics[width=0.6\textwidth,height=0.6\linewidth]{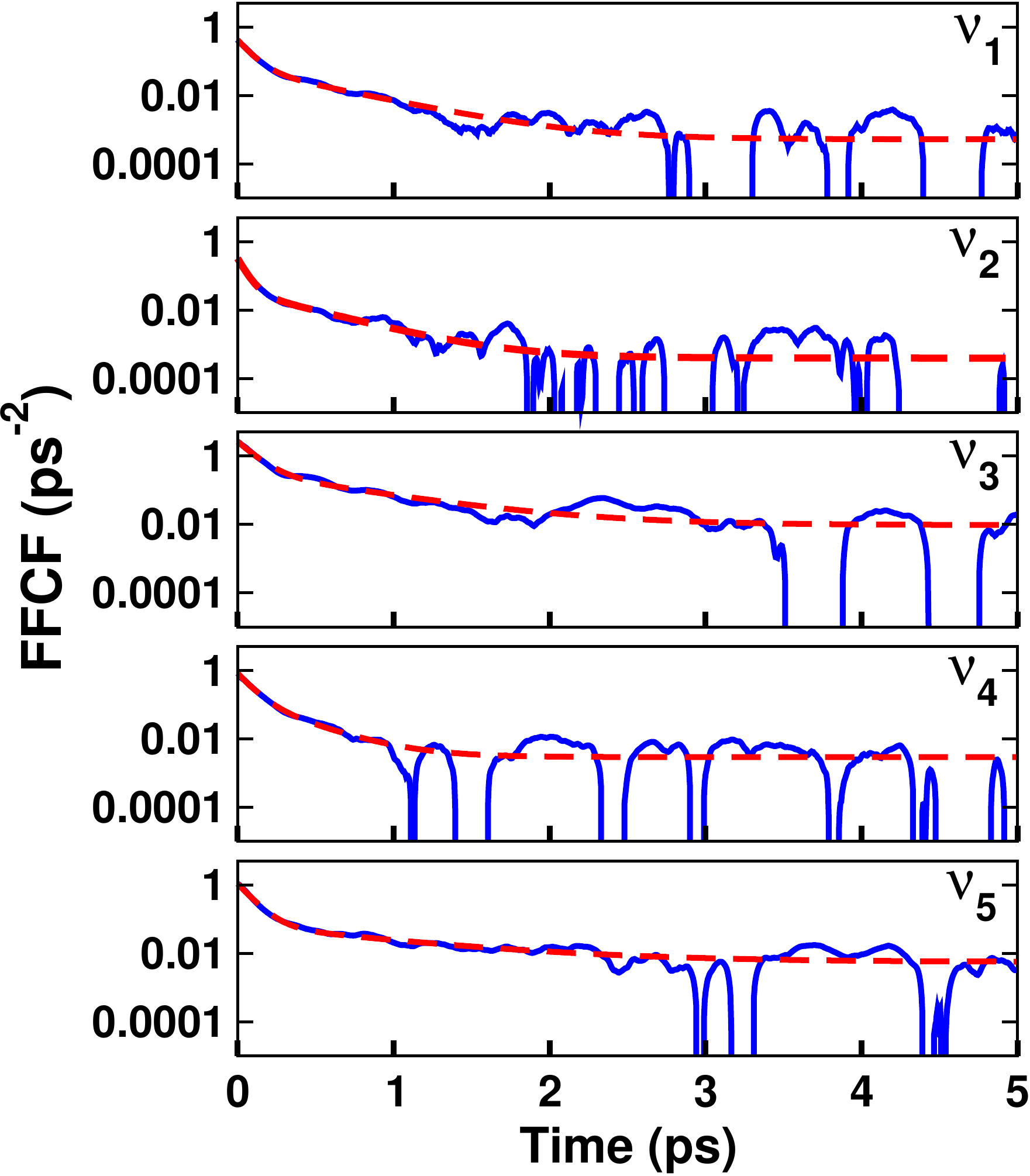}
\caption{FFCFs from INM frequency calculations for F-PhOH in water
  from a 5 ns simulation using MTP. The FFCFs for the 5 modes $\nu_1$
  to $\nu_5$ between 1100 and 1400 cm$^{-1}$ are reported. The solid
  lines are the raw FFCF data and the dashed lines show the
  corresponding fit with fitting parameters reported in Table
  S5. Logarithmic scale is chosen for the $y$-axis.}
\label{sifig:ffcf}
\end{center}
\end{figure}

\begin{figure}[H]
\begin{center}
\includegraphics[width=0.7\textwidth]{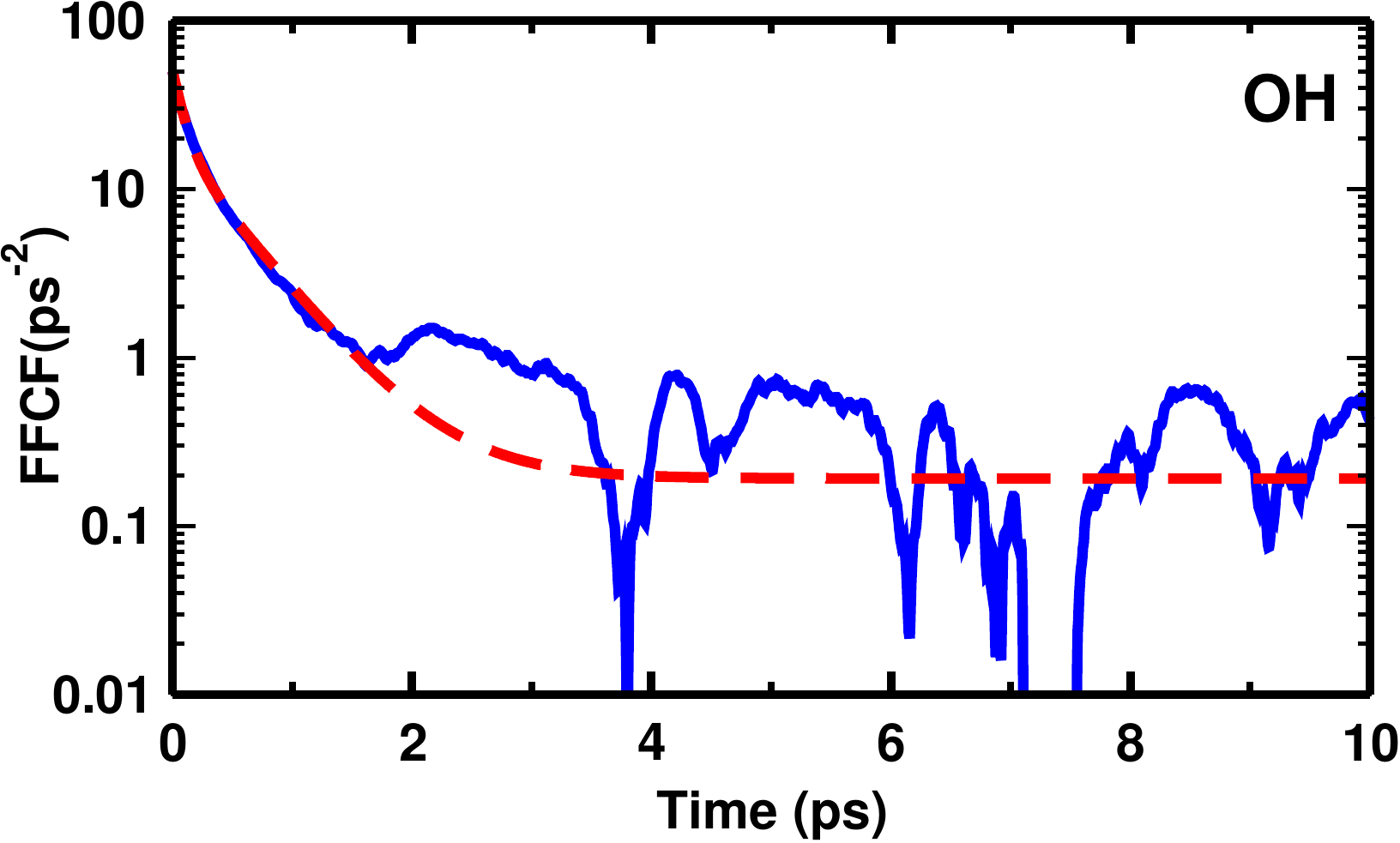}
\caption{FFCFs from INM frequency calculations for F-PhOH in water
  from a 5 ns simulation using MTP. The FFCFs for the OH mode is
  reported. The solid lines are the raw FFCF data and the dashed lines
  show the corresponding fit with fitting parameters reported in Table
  2. Logarithmic scale is chosen for the $y$-axis.}
\label{sifig:ffcfoh}
\end{center}
\end{figure}

\bibliography{refs}